\documentclass[twoside, a4paper]{article}

\let\UnmodifSec=\section
\renewcommand{\section}{\setcounter{equation}{0}\UnmodifSec}

\def \vhi{\varphi}

\def\Im{{\rm Im\,}}
\def\bR{{\bf R}}
\def\bN{{\bf N}}

\def\bC{{\bf C}}
\def\bCp{{\bC_+}}
\def\bCm{{\bC_-}}
\def\bCpm{{\bC_\pm}}

\def\bbCp{\ovl{\bC_+}}
\def\bbCm{\ovl{\bC_-}}
\def\bbCpm{\ovl{\bC_\pm}}

\def\CC{{\cal C}}

\def\QQ{{\cal Q}}
\def\RR{{\cal R}}

\def\VV{{\cal V}}

\def\wt{\widetilde}
\def\wh{\widehat}

\def\ovl{\overline}
\def\unl{\underline}
\def\interior#1{\setbox1=\hbox{$#1$}\rlap{$#1$}\kern0.4\wd1\raise1.1\ht1%
\hbox{$\scriptstyle \circ$}}

\def\chu{{\check u}}
\def\hphi{\phi}
\def \ghi{g}
\newenvironment{acknowledgement}{
\vskip 0.25 truecm
\noindent {\bf Acknowledgement.\ }}{}

\newtheorem{lemma}{Lemma}

\newtheorem{theorem}{Theorem}
\newtheorem{remark}{Remark}[section]

\newcommand{\PROOF}{\noindent{\it Proof.\ \ }}

\renewcommand{\arraystretch}{1.5}
\title{Existence and properties of $p$-tupling fixed points}
\author{Henri Epstein%
\\
Institut des Hautes Etudes Scientifiques,\\
91440 Bures-sur-Yvette, France}

\date{March 2000}

\begin{document}
\maketitle

\begin{abstract}
We prove the existence of fixed points of $p$-tupling renormalization
operators for interval and circle mappings having a critical point of
arbitrary real degree $r>1$. Some properties of the resulting maps
are studied: analyticity, univalence, behavior as $r$ tends to
infinity.
\end{abstract}

\section{Introduction}

\label{intro}
Two problems have a strong 
resemblance, and have both found their origin
in the theory of period doubling for maps of the interval \cite{F1,F2,CT}.
The first is to prove the existence and properties of solutions
of the $(p+1)$-Cvitanovi\'c-Feigenbaum functional equation, i.e.  
fixed points of the $(p+1)$-tupling operator $\RR_{p+1}$:
\begin{equation}
\ghi(x) = (\RR_{p+1}\ghi)(x) =
- {1\over\lambda } \ghi^{p+1}(-\lambda x),\ \ \ \ 
\ghi(0)=1.
\label{1.1}\end{equation}
Here $\ghi$ is required to be an even, $\CC^1$ map of $[-1,\ 1]$ into itself, 
strictly decreasing on $[0,\ 1]$ and 
$\lambda = -\ghi^{p+1}(0)$ is required to be in $(0,\ 1)$. More precisely, 
the restrictions $\ghi_+$ and $\ghi_-$ to 
$[0,\ 1]$ and $[-1,\ 0]$, respectively, must satisfy
\begin{equation}
\ghi_+ = - {1\over\lambda } \ghi_-^{p-1}\circ\ghi_+^2\circ (\lambda),\ \ \ \ 
\ghi(0)=1.
\label{1.2}\end{equation}
Denoting $u$ the inverse function of $\ghi_+$, and 
$\check u(z) = u(-z)$, this can be reexpressed as
\begin{equation}
u = {1\over\lambda} u\circ \check u^p \circ \lambda\ .
\label{1.3}\end{equation}
We shall also require $\ghi_+$ to have the form 
\begin{equation}
\ghi_+(x) = f(x^r) \ \ \ \forall x \in [0,\ 1],
\label{1.4}\end{equation}
where $r > 1$ is a real number, and $f$ is real-analytic 
on $[0,\ 1]$, with $f'(x) <0$ on this closed interval. 
The class of those $\ghi$ having the property (\ref{1.4}) for a fixed $r$ 
is left invariant by $\RR_{p+1}$. 

The second problem is to prove the existence and properties of solutions
of the system 
\begin{equation}
\begin{array}{ccc}
\displaystyle
\eta &=& -{1\over\lambda} \eta^p \circ \xi \circ (-\lambda)\ ,\\
\displaystyle
\xi &=& -{1\over\lambda} \eta \circ (-\lambda),\ \ \ \xi(0) = 1\ ,
\end{array}
\ \ \ \ \ \ \lambda = -\eta(0) \in (0,\ 1).
\label{1.5}\end{equation}
Here $\xi$ is a real $\CC^1$, strictly increasing function defined on 
a certain interval $[-L,\ 0]$ of the negative real axis ($L >1$) and 
satisfies $\xi(x) > x$ on this interval. 
Again $\xi$ is required to be of the form
\begin{equation}
\xi(x) = f(|x|^r) \ \ \ \forall x \in [-L,\ 0]\ ,
\label{1.6}\end{equation}
where $r > 1$ is a real number, and 
$f$ is real analytic without critical points on $[0,\ L^r]$. 
Let $-u$ be the inverse function of $\xi$, and $\check u(z) = u(-z)$. 
Then (\ref{1.5}) implies
\begin{equation}
u = {1 \over \lambda^2} u\circ \lambda \circ \check u^p \circ \lambda\ .
\label{1.7}\end{equation}
The system (\ref{1.5}) is part of the theory, initiated in \cite{FKS} and 
\cite{ORSS},
of critical circle mappings whose rotation number has 
the continued fraction expansion $[p,\ p,\ \dots, p,\ \dots]$. 
It is natural to attempt a unified treatment of the two functional equations 
(\ref{1.3}) and (\ref{1.7}) by introducing an interpolating parameter 
$\nu \in [1,\ 2]$ and considering the functional equation 
\begin{equation}
u = {1 \over \lambda^\nu} u\circ \lambda^{\nu-1} 
\circ \check u^p \circ \lambda\ .
\label{1.8}\end{equation}
As a device for avoiding repetitions, this works rather well  
for $p=1$, (\cite{EE,E2,E3}). It is much less effective, 
as we shall see, for $p > 1$.
It is also of some interest to consider the case when $\nu <1$.
The history of this subject is long, even if restricted to
rigorous results (see e.g. \cite{L1,L2}), 
and the literature has experienced a veritable
explosion in recent times. For the case of interval maps, the paper
of M. Lyubich \cite{Ly} (a kind of culminating point)
contains a historical note and references to which I refer the reader.
For the case of circle maps, the reader is referred to the paper of
M. Yampolsky \cite{Y} and to references therein. 
However the literature has tended to concentrate on the case of
integer $r$, with notable exceptions such as \cite{CEL,JR,M,MO}. Another,
most important exception is the whole theory of ``real a priori bounds''
(see \cite{dMvS,S1,S2,Sw,dFdM} and other references given in \cite{Ly,Y}).
In this paper,
we look for solutions of the functional equations (\ref{1.8}), for arbitrary
real $r$, which are subjected 
to some additional constraints (see Section 3). 
All the available theoretical and numerical evidence indicates that, 
for each $\nu \in [1,\ 2]$, each $r > 1$, and each $p \ge 1$, there is 
one and only one solution obeying all the constraints. This 
suggests that the solution (and in particular $\lambda$)
must depend analytically on the 
parameters $\nu$, $r$, and $p$.
For $p=1$, it has been proved in \cite{E3}
that solutions exist for all $\nu \in [1,\ 2]$ and all $r >1$, and the
proof extends without any change to the case $\nu \in (0,\ 1]$ provided
$r\nu >1$. In the case $\nu = 1$, the existence of solutions for all
$p$ and all $r>1$ has been proved by M.~Martens \cite{M}, whose results
go much farther since they include all possible periodic points and
kneading sequences.
In this paper, the existence 
of solutions will be proved, by another method, 
in the case $1 \le \nu \le 2$, for all $r > 1$ and all
(integer) $p \ge 1$.
It will be seen that in the case $0 < \nu \le 1$, 
the condition $r\nu - 1-(p-1)(1-\nu) >0$ is necessary and sufficient
for the existence of solutions. This work had remained unfinished for a
long time\footnote{The case $\nu = 1$, all $r$ and $p$, was
presented at the Meeting on new developments in Mathematical Physics 
and Neuroscience 
(Hunziker-Hepp Fest), ETH, Zurich, 21-23/9/1995.}
when I belatedly became aware of the paper of 
B.~Mestel and A.~Osbaldestin \cite{MO}, devoted to the proof of
the existence of a period 2 point of the doubling operator for
non-even maps (with arbitrary $r>1$). One of the ideas in that paper
allowed me to finish the proof of existence in the case $\nu >1$
(see Subsection \ref{opN}).

Section 2 collects some notations and well-known or straightforward facts
(see \cite{D,V}). 
Sections 3-6 contain the proofs of existence.
In Section 7 some properties of the solutions are
derived (univalence, boundedness). In Section 8 it is shown that
for $\nu \in (0,\ 1]$, when $r$ tends
to infinity the solutions behave similarly to those of the case 
$p=1$ (see \cite{EW,E1,EE}).

\begin{acknowledgement}
I wish to thank Oscar Lanford and Michael Yampolsky for many
helpful discussions. I am also indebted to O.~Lanford for his kind
permission to include the contents of Subsection \ref{commut}.
\end{acknowledgement}
\section{Notations and preliminaries}

\label{prel}
\renewcommand{\arraystretch}{1.5}
1. We denote
${\bf C}_+ = - {\bf C}_- =
\{z \in {\bf C} :\ {\rm Im}\ z >0 \}$. A function $f$
is a Herglotz or Pick function [D] (and $-f$ is an anti-Herglotz 
function) if it is holomorphic in 
${\bf C}_+ \cup {\bf C}_-$, $f(z^*)= f(z)^*$, and
$f$ maps $\bC_+$ (resp. $\bC_-$) into its closure, 
$f({\bf C}_{\pm}) \subset \overline {\bC_{\pm}}$.
If $f$ is also holomorphic on a real non-empty
open segment $(a,\ b)$, then, for each $x \in (a,\ b)$, and
each $N\in \bN$ the $N\times N$ matrix $M$ with components
$M_{jk} = D^{j+k+1}f(x)/(j+k+1)!$, ($0 \le j,\ k < N$), is positive. 
This follows immediately from the Herglotz 
integral representation theorem ([D], pp. 20 ff.).
The case $N=2$
shows that if $f$ is not a constant, then
$f'(x) >0$ $\forall x \in (a,\ b)$, and $f$ has non-negative
Schwarzian derivative $Sf = (f''/f')'- (f''/f')^2/2$ in $(a,\ b)$.
Denote $v = f''/f'$ and suppose $a <x < y< b$. If
$v$ does not vanish in $[x,\ y]$,
\begin{equation}
{1 \over v(x)} - {1 \over v(y)} \ge {y-x \over 2}.
\end{equation}
If $v(x) >0$ then $v(y) >0$ since $v$ is increasing, hence
$v(x) \le 2/(y-x)$, which also holds if $v(x) \le 0$.
Similarly $v(y) \ge -2/(y-x)$. Letting $y$ tend to $b$ or 
$x$ tend to $a$, we find:
\begin{equation}
-{2 \over z-a} \le {f''(z)\over f'(z)} \le {2 \over b-z}\ \ \ \ 
\forall z \in (a,\ b).
\label{AA.0}\end{equation}
If $f((a,\ b))$ has a finite upper bound, then $f|(a,\ b)$ extends
continuously to $(a,\ b]$ with $f(b) = \sup f((a,\ b))$, and similarly
if there is a finite lower bound. If $f$ maps $[a,\ b]$
into $(a,\ b)$, it has a fixed point in $(a,\ b)$ which
(by Schwarz's lemma) is unique and attractive;
in this case every subinterval of $(a,\ b)$ which contains the
fixed point is mapped into itself by $f$. If $F$ is an increasing 
function with non-negative Schwarzian on $(0,\ +\infty)$ 
(in particular if $F$ is a Herglotz function holomorphic in 
$\bC_+\cup\bC_-\cup(0,\ +\infty)$), then its restriction
to $(0,\ +\infty)$ is concave as a special case of (\ref{AA.0}).
The following corollary will be needed later:

\begin{lemma} \label{noses}
Let $A < a < b < B$ be real numbers, and $f$ be a Herglotz function
holomorphic in $\bC_+\cup\bC_-\cup (A,\ B)$. Then, for each
$z \in (a,\ b)$,
\begin{equation}
f(z) \ge {(B-b)(z-a)f(b) + (B-a)(b-z)f(a) \over (b-a)(B-z)}\ .
\label{AA.1}\end{equation}
\end{lemma}

\noindent{\it Proof.}
We define $F(z) = f(B-z^{-1})$, i.e.
$f(z) = F(1/(B-z))$. Then $F$ is a 
Herglotz function holomorphic in $\bC_+\cup\bC_-\cup (1/(B-A),\ +\infty)$.
Setting now $a'= 1/(B-a)$, $b' = 1/(B-b)$ and $z' = 1/(B-z)$, with
$z \in (a,\ b)$, the concavity of $F$ implies
\begin{equation}
F(z') \ge {z'-a' \over b'-a'} F(b') + {b'-z' \over b'-a'} F(a')\ ,
\end{equation}
which translates back into (\ref{AA.1}).

\noindent 2. Let $A$, $B$, $A'$, $B'$ be strictly positive real numbers.
Then the homographic function 
\begin{equation}
z \mapsto m(z\ ;\ A, B, A', B') = 
{z \left ( {1 \over A} + {1 \over B} \right ) \over
z \left ( {1 \over AB'} - {1 \over BA'} \right ) +
{1 \over A'} + {1 \over B'}}
\label{2.1}\end{equation}
is a bijection of $\bC_+ \cup \bC_- \cup [-A,\ B]$ onto 
$\bC_+ \cup \bC_- \cup [-A',\ B']$, and fixes $0$. Its derivative at 
$0$ is
\begin{equation}
m'(0\ ;\ A, B, A', B') = 
{A'B'(A+B) \over AB(A'+B')}\ .
\label{2.2}\end{equation}
Hence if $f$ is a holomorphic map of $\bC_+ \cup \bC_- \cup (-A,\ B)$ 
into $\bC_+ \cup \bC_- \cup (-A',\ B')$ which fixes $0$, it follows from 
Schwarz's lemma applied to $m^{-1}\circ f$, that
$|f'(0)| \le m'(0\ ;\ A, B, A', B')$.

\noindent 3. Let $b,\ s$ be real numbers,
with $0< b < 1$, and $s >1$. Then the homographic function
\begin{equation}
z \mapsto \chi_{b,s}(z) = 1 +
b^{s-1}\,{b(1+b)(z-1) \over 1+b-b^2(z-1)}\ 
\label{2.2.1}\end{equation}
is holomorphic in $\Omega(-1/b,\ 1/b^2)$, Herglotzian, and
\begin{equation}
\chi_{b,s}(-1/b) > 0,\ \ \
\chi_{b,s}(1/b^2) < 1/b^2,\ \ \
\chi_{b,s}(1) = 1,\ \ \ \chi'_{b,s}(1) = b^s\ .
\label{2.2.2}\end{equation}

\noindent 4. In the sequel, if $s$ and $t>s$ are real numbers,
we shall denote $\Omega(s,\ t)$ the domain 
\begin{equation}
\Omega(s,\ t) = \bC_+ \cup \bC_- \cup (s,\ t).
\label{2.3}\end{equation} 
If $u_-$ and $u_+$ are two real numbers in $(0,\ 1)$, we denote 
${\bf E}_0(u_-,\ u_+)$ the space of functions $\psi$, holomorphic
and anti-Herglotzian in $\Omega(-1/u_-,\ 1/u_+)$,
and such that $\psi(0) =1$, $\psi(1)=0$.
Such a function has an integral representation 
\begin{equation}
\log \psi(z) =
\int_{\bR \setminus (-1/u_-,\ 1)}\ 
\sigma(t)\left [ {1\over t} - {1\over t-z} \right ]dt\ ,
\ \ \ \forall z \in \Omega(-1/u_-,\ 1)\ .
\label{2.4}\end{equation}
Here $\sigma$ is an $L^\infty$ function with $0\le\sigma\le 1$ and 
$\sigma(t) = 1$ for all $t \in [1,\ 1/u_+]$. It follows that $\psi$
satisfies the following inequalities:
\begin{equation}
\begin{array}{ccc}
\displaystyle
{\psi(z)(1-u_+z) \over 1-z} &\le 1 \le
\displaystyle
{\psi(z)(1+u_-z) \over 1-z} \ \ \ 
&{\rm for\ all}\ \ z \in (0,\ 1/u_+) \setminus \{1\},\ \ \\
{ }&{ }&{\rm reversed\ for}\ z \in (-1/u_-,\ 0),
\end{array}
\label{2.5}\end{equation}
\begin{equation}
\begin{array}{ccl}
\displaystyle
{1-u_+ \over (1-z)(1-u_+z)} &\le\ \ 
\displaystyle
-{\psi'(z) \over \psi(z)} &\le
\displaystyle
{1+u_- \over (1-z)(1+u_-z)}\\
&{}&{\rm for\ all}\ \ z \in (-1/u_-,\ 1/u_+) \setminus \{1\} .
\end{array}
\label{2.6}\end{equation} 
Suppose now that $\psi \in {\bf E}_0(u_-,\ u_+)$ has a finite upper 
bound $M$ on $(u_-,\ u_+)$. (By (\ref{2.5}), $M$ must satisfy 
$M \ge \psi(-1/u_-) \ge (1+u_-)/(u_+ + u_-)$ .) Then
\begin{equation}
\int_{\bR \setminus (-1/u_-,\ 1)}\ 
{\sigma(t)\,dt \over t(1+u_-t)} \ \ \le \ \ \log M\ ,
\end{equation}
Therefore, if $-1/u_- <z <0$,
\begin{eqnarray}
-{\displaystyle \psi'(z) \over \displaystyle \psi(z)} 
&=& \displaystyle \int_{ \bR \setminus (-1/u_-,\ 1)}\ 
{\sigma(t) \over \displaystyle t(1+u_-t)}\ 
{\displaystyle t(1+u_-t) \over \displaystyle (t-z)^2}\,dt
\nonumber \\ 
&\le&
(\log M)\,\displaystyle \max_{t \not\in (-1/u_-,\ 1)}
{\displaystyle t(1+u_-t) \over \displaystyle (t-z)^2} \ \ \le\ \ 
{\displaystyle \log\,M \over \displaystyle (-4z)(1+u_-z)}\ .
\label{2.7}\end{eqnarray}

\section{More precise statement of the problem}

\label{state}
We begin with a few heuristic considerations. 
It is easy to see that if $u$ is a solution of (\ref{1.8}), it will
analytically extend to the real interval $(-1/\lambda,\ 1)$. Moreover,
since the function $f$ is analytic without critical point in an open
real interval containing 0, its inverse function, denoted $U$, will
be analytic, with strictly negative derivative, in an open interval
containing 1. The functions $u$ and $U$ are related by
$U(z) = u(z)^r$. Thus
\begin{equation}
U(z) = {1\over \lambda^{r\nu}}
U(\lambda^{\nu-1}\check u^p(\lambda z))
\label{3.1}\end{equation}
should hold wherever both sides are analytic.
The main condition which we impose on the solutions we seek is that
$u$ and $U$ be anti-Herglotz functions. It is in fact sufficient to
impose this condition on $u$. Indeed denote
\begin{equation}
\vhi = \lambda^{\nu -1}\check u^p \circ \lambda\ .
\label{3.2}\end{equation}
Then $\vhi$ is Herglotzian and the equation (\ref{3.1}), rewritten as
\begin{equation}
U(z) = {1\over \lambda^{r\nu}} U(\vhi(z))
\label{3.3}\end{equation}
shows first that $\vhi(1)$ must be a zero of $U$, i.e. that
$\vhi(1) =1$, and then that $U$ is a linearizer of $\vhi$ at 1,
the multiplier $\vhi'(1)$ being equal to $\lambda^{r\nu} < 1$.
Therefore $U$ is also anti-Herglotzian, and is holomorphic in
the basin of attraction of 1 for $\vhi$. The reasons for imposing
the Herglotz condition have been given e.g. in \cite{EE,E2}.
It is more convenient to work with
$\psi = U/U(0)$ rather than with $U$, and we denote
$z_1/\lambda^{\nu-1}$ the quantity $u(0)$. 
This implies $z_1 = \lambda^{\nu -1}\chu(0) \le \vhi(0) < 1$.
We thus adopt the following definition:

Given two real numbers $\nu \in (0,\ 2]$, $r>1$, and an integer $p \ge 1$,
a solution associated with these values consists of two functions
$\psi$ and $u$ and two real numbers $\lambda \in (0,\ 1)$ and
$z_1 \in (0,\ 1)$ with the following properties:

\begin{description}
\item{(1)} $\psi$ is an anti-Herglotzian function holomorphic in
$\Omega(-1/\lambda,\ 1/a)$ for some $a\in (0,\ 1)$ with
$\psi(1) = 0$ and $\psi(0) = 1$.

\item{(2)} $u$ is holomorphic and anti-Herglotzian
in $\Omega(-1/\lambda,\ 1)$, and is given there by
\begin{equation}
u(z) = {z_1 \over \lambda^{\nu-1}}\, \psi(z)^{1/r}\ .
\label{3.4}\end{equation}

\item{(3)} The identity
\begin{equation}
\psi(z) = {1\over\lambda^{r\nu}}
\psi(\lambda^{\nu-1}\check u^p(\lambda z))
\label{3.5}\end{equation}
holds for all $z \in \Omega(-1/\lambda,\ 1/a)$, where again
$\check u(z) = u(-z)$.
\end{description}

\noindent The following theorem will be proved.

\begin{theorem}
\label{exist}
(i) For any integer $p \ge 1$ and any real $\nu \in (0,\ 1]$,
there exist solutions if and only if $r$ satisfies
\begin{equation}
r\nu - 1-(p-1)(1-\nu) > 0\ .
\label{3.5.1}\end{equation}
(ii) For any integer $p \ge 1$ and any real $\nu \in [1,\ 2]$,
there exist solutions for all real $r>1$.
\end{theorem}

The necessity of the condition (\ref{3.5.1}) will be shown in 
the next section, but it is easy to see that the 
conditions we have imposed
require $r\nu >1$. It suffices to consider the case 
$0< \nu \le 1$. In this case, we must have
\begin{equation}
\psi(-1/\lambda) = {1 \over \lambda^{r\nu}}
\psi(\vhi(-1/\lambda)) \le {1 \over \lambda^{r\nu}}
\ \ \Rightarrow\ \ 
u(-1/\lambda) \le {z_1 \over \lambda^{2\nu -1}} < {1 \over \lambda}\ ,
\label{3.6}\end{equation}
from which it follows that $\vhi = \lambda^{\nu-1} \check u^p \circ \lambda$
is analytic in $\Omega(-1/\lambda,\ 1/\lambda^2)$ and that:
\begin{equation}
\vhi(-1/\lambda) \ge 0,\ \ \ 
\vhi(1) = 1,\ \ \ 
\vhi(1/\lambda^2) \le {z_1 \over \lambda^\nu}\ .
\label{3.7}\end{equation}
We can apply Remark 2 of Section \ref{prel}, i.e. Schwarz's lemma as in
(\ref{2.2}) to bound $\vhi'(1)$, with
$A = 1+1/\lambda$, $B = 1/\lambda^2 - 1$, 
$A' = 1$, $B' = \lambda^{-\nu} -1$ and find
\begin{equation}
\vhi'(1)  = \lambda^{r\nu} \le  
{\lambda\,(1- \lambda^\nu) \over (1- \lambda^2)}\ \le\ 
{\lambda \over 1+\lambda}\ .
\label{3.8}\end{equation}
This implies $\lambda^{r\nu-1} <1$, i.e. $r\nu > 1$.

In the case $\nu > 1$, it is well-known (see \cite{JR}), and easy to verify, 
that for $r=1$, $p \ge 1$, there is a
solution such that $\psi(z) = 1-z$, all functions $u$, $\vhi$, etc.
are affine, $z_1 = \lambda^{\nu-1}$,
and $\lambda$ is the unique solution in $(0,\ 1)$ of
\begin{equation}
\lambda^{\nu} + p \lambda^{\nu -1} -1 = 0\ .
\label{3.9}\end{equation}
Moreover it is proved in \cite{JR} that (in the case $p = 1$, $\nu = 2$)
there exist solutions for all sufficiently small $r-1 > 0$.

The proof of Theorem \ref{exist} will occupy Sections 
\ref{crpnu}-\ref{p>1nu>1}. Many repetitions occur in these sections,
since variations of the same method apply to several cases.
But avoiding the repetitions would produce more obscurity than
brevity.

\section{Existence for $r>1$, $p\ge 1$ and $\nu \in (0,\ 1]$}

\label{crpnu}
In this section, $r > 1$ and $\nu \in (0,\ 1]$ are fixed real numbers
such that $r\nu > 1$, and $p \ge 1$ is a fixed integer. 
The real number $b \in (0,\ 1)$ is also fixed, but its value
will be chosen later (as a function of $r$). For any 
two $s,\ t \in (0,\ 1)$, we denote
\begin{equation}
h_{s,t}(z) =
{z(s+1) \over z(s-t) +1+t}\ \ \ \
\left \{\ \matrix{
0 &\mapsto &0\cr
1 &\mapsto &1\cr
-1/s &\mapsto &-1/t\cr} \right.
\label{4.1}\end{equation}
Obviously 
$h_{s,t} = h_{t,s}^{-1}$.

We denote $\QQ_0(b,\ r\nu)$ the space of all functions $\Phi$ with the
following properties:

\begin{description}
\item{(Q1)}\quad  $\Phi$ is a Pick function holomorphic in the domain:
\begin{equation}
\Omega(-1/b,\ 1/b^2) = \bC_+ \cup \bC_- \cup
\left (-{1\over b},\ {1\over b^2} \right )\ ,
\label{4.3}\end{equation}
and maps this domain into itself.

\item{(Q2)}\quad  $\Phi(z) \ge 0$ for all $z \in (-1/b,\ 1/b^2)$,

\item{(Q3)}\quad  $\Phi(1) = 1$, and $0< \Phi'(1) \le b^{r\nu}$. 
\end{description}

We regard  $\QQ_0(b,\ r\nu)$ as a subset of the real Fr\'echet space of the
self-conjugated functions holomorphic in $\Omega(-1/b,\ 1/b^2)$, 
equipped with the topology of uniform convergence on compact subsets.
We shall define a continuous
operator $B(b,\ r,\ p,\ \nu)$ on this space by describing its action on an
arbitrary $\Phi_0 \in \QQ_0(b,\ r\nu)$.

Given $\Phi_0 \in \QQ_0(b,\ r\nu)$, we denote 
$\lambda = \Phi'_0(1)^{1/r\nu}$. By (Q3), $0<\lambda \le b$. 
We define a function $\vhi_0$ by
\begin{equation}
\vhi_0 = h_{b,\lambda}\circ\Phi_0\circ h_{b,\lambda}^{-1}\ .
\label{4.5}\end{equation}
If $\lambda = b$, $h_{b,\lambda}$ is the identity. Otherwise, 
since $\lambda < b$, its pole is below $-1/b$
and $h_{b,\lambda}$ maps $\Omega(-1/b,\ 1/b^2)$ onto
$\Omega(-1/\lambda,\ 1/a_0(\lambda))$ where
\begin{displaymath}
{1 \over a_0(\lambda)} = h_{b,\lambda}\left({1\over b^2}\right )\ ,
\end{displaymath}
\begin{equation}
b^2\ \  \le\ \ a_0(\lambda) = 
b - \lambda(1-b)\ \ <\ \ b.
\label{4.6}\end{equation}

The function $\vhi_0$ possesses the following properties:

\begin{description}
\item{(Q$'1$)}\quad  $\vhi_0$ is a Pick function holomorphic in
$\Omega(-1/\lambda, 1/a_0(\lambda))$, and maps this domain into itself.
\item{(Q$'2$)}\quad  $\vhi_0(z) \ge 0$ for all 
$z \in (-1/\lambda,\ 1/a_0(\lambda))$.
\item{(Q$'3$)}\quad  $\vhi_0(1) = 1$, and $\vhi'_0(1) = \lambda^{r\nu}$.
\end{description}

We denote $\psi$ the linearizer of $\vhi_0$ normalized by the 
condition $\psi(0) = 1$, i.e. the unique function, holomorphic in 
$\Omega(-1/\lambda, 1/a_0(\lambda))$, such that 
\begin{equation}
\psi(z) = {1\over \lambda^{r\nu}} \psi(\vhi_0(z)) \ \ \ 
\forall z \in \Omega(-1/\lambda, 1/a_0(\lambda)),\ \ \ 
\psi(1) = 0,\ \ \ \psi(0) = 1\ .
\label{4.7}\end{equation}
This is an anti-Herglotz function given, as it is well known ([Mi, V]), by
\begin{equation}
\psi(z) = h(z)/h(0),\ \ \
h(z) = \lim_{n\rightarrow \infty} {1\over \lambda^{nr\nu}}
(\vhi_0^n(z)-1)\ .
\label{4.8}\end{equation}
The limit converges uniformly on compact subsets of 
$\Omega(-1/\lambda,\ 1/a_0(\lambda))$, which is a 
basin of attraction of 1 for $\vhi_0$. It is easy to check that $\psi$
depends continuously on $\vhi_0$. On $[-1/\lambda,\ 1/a_0(\lambda))$, 
$\psi$ is strictly decreasing and, because $0 \le \vhi_0(-1/\lambda))<1$, 
\begin{equation}
\psi(-1/\lambda) = {1 \over \lambda^{r\nu}} \psi(\vhi_0(-1/\lambda))\ \ 
\le {1 \over \lambda^{r\nu}}\ .
\label{4.9}\end{equation}
$\psi$  satisfies the inequalities (\ref{2.5}) and (\ref{2.6}), 
with $u_- = \lambda$ and $u_+ = a_0(\lambda)$.
We also define
\begin{equation}
v(z) = (\psi(-z))^{1/r} \ \ \ \forall 
z \in \bC_+ \cup \bC_- \cup (-1,\ 1/\lambda)\ . 
\label{4.10}\end{equation}
$v$ is a Pick function which extends to a strictly increasing 
continuous function on $[-1,\ 1/\lambda]$. It satisfies:
\begin{equation}
v(-1) = 0,\ \ \ v(0) = 1,\ \ \ 
v(1/\lambda) \le \lambda^{-\nu}\ .
\label{4.11}\end{equation}

We now show that there is a unique $z_1 \in (0,\ 1)$ such that
\begin{equation}
\left ( z_1 \lambda^{1-\nu}v \right )^p (\lambda) = 
\lambda^{1-\nu}\ .
\label{4.12}\end{equation}
As a consequence of the inequality in (\ref{4.11}), the function 
$(s\lambda^{1-\nu}\,v)^k$ is defined on $[-1,\ 1/\lambda]$
for every $s \in [0,\ 1]$ and every integer $k \ge 0$. The functions
$s \mapsto x_k(s) = (s\lambda^{1-\nu}\,v)^k(\lambda)$ are thus defined, 
continuous, and strictly increasing in $s$ for $k >0$ and $s \in [0,\ 1]$.
For $s_* = \lambda^{\nu}/v(\lambda)$, $x_k(s_*) = \lambda$ for all $k$.
For $s > s_*$, $x_1(s) > x_1(s_*) = \lambda = x_0(s)$, hence
$x_0(s) < x_1(s) < \dots x_{p+1}(s)$. Since 
$x_p(1) > x_1(1) = \lambda^{1-\nu} v(\lambda) > \lambda^{1-\nu}$,
there exists a unique 
$z_1 \in (s_*,\ 1)$ such that $x_p(z_1) = \lambda^{1-\nu}$. 
In the case $p=1$, (\ref{4.12}) reduces to $z_1 = 1/v(\lambda)$.

The derivative $x'_p(z_1)$ is strictly
positive. Therefore, 
if $v$ is allowed to change slightly, $z_1$ can be computed by a
Cauchy integral along a small circle which remains fixed. Thus
$z_1$ depends continuously (in fact analytically) on $v$, hence on $\vhi_0$.

We denote $\zeta_j = (z_1 \lambda^{1-\nu} v)^j (\lambda)$, 
$(0 \le j \le p+1)$. By the preceding argument,
\begin{equation}
\lambda = \zeta_0 < \zeta_1 < \dots < \zeta_p = \lambda^{1-\nu} < 
\zeta_{p+1} = z_1 \lambda^{1-\nu} v(\lambda^{1-\nu}),
\label{4.13}\end{equation}
Since $v(\lambda^{1-\nu}) \le \lambda^{-\nu}$,
\begin{equation}
z_1  > \lambda^\nu\ ,\ \ \ \ z_1 \lambda^{1-\nu} > \lambda\ .
\label{4.14}\end{equation}
Note also that, since $v(\lambda) > v(0) = 1$,
\begin{equation}
\zeta_1 > z_1 \lambda^{1-\nu}\ .
\label{4.15}\end{equation} 
The last inequality in (\ref{4.13}), the upper bound on 
$\psi(-\lambda^{1-\nu})$ from (\ref{2.5}), and $\nu \le 1$ give
\begin{equation}
z_1 \ \ \ge \ \ 
\left ( {1 - \lambda^{2-\nu} \over 1 + \lambda^{1-\nu}} \right )^{1\over r}
\ \ > \ \ 
{1 - \lambda \over 2}
\label{4.17}\end{equation}
We can now define a new function $\vhi$ by:
\begin{equation}
\vhi(z) = \lambda^{\nu-1}
\left ( z_1 \lambda^{1-\nu} v \right )^p (\lambda z)\ ,
\ \ \ z\in \Omega(-1/\lambda,\ 1/\lambda^2)\ .
\label{4.18}\end{equation}
In this domain, $\vhi$ is a Pick function, which extends continuously
to the ends of its real interval of definition, and
\begin{equation}
\vhi(-1/\lambda)\  = 0\ \ \ {\rm if}\ \ p = 1,
\label{4.19.a}\end{equation}
\begin{equation}
\vhi(-1/\lambda)\  =\  
\lambda^{\nu-1} ( z_1 \lambda^{1-\nu} v )^{p-1}(0)
\ \ge \ z_1 \ >\ \lambda^\nu\ \ \ {\rm if}\ \ p \ge 2 ,
\label{4.19.b}\end{equation}
\begin{equation}
\vhi(1)\ =\ 1\ ,\ \ \ \ 
\vhi(1/\lambda^2) = \lambda^{\nu-1}(z_1 \lambda^{1-\nu} v)^p(1/\lambda) 
\le\  z_1/\lambda^\nu \ .
\label{4.19.c}\end{equation}
The domain $\Omega(-1/\lambda,\ 1/\lambda^2)$ is thus a basin of
attraction of the fixed point 1 of $\vhi$. 
This domain contains $\Omega(-1/\lambda,\ 1/a_0(\lambda))$ since
$a_0(\lambda) \ge \lambda^2$ (see (\ref{4.6})).

We now use Schwarz's lemma, as mentioned in Section \ref{prel}, to
obtain an upper bound for $\vhi'(1)$. If $p \ge 2$, 
\begin{displaymath}
\vhi'(1) \le {A'B'(A+B)\over AB(A'+B')}\ \ {\rm with}
\end{displaymath}
\begin{displaymath}
A = 1+{1\over\lambda},\ \ B = {1 \over\lambda^2}-1\ ,\ \ 
A'= 1- z_1,\ \ \ B' = {z_1 \over \lambda^\nu} -1\ .
\end{displaymath}
This gives
\begin{equation}
\vhi'(1) \le {\lambda (1-z_1)(z_1 - \lambda^\nu) \over
z_1(1-\lambda^\nu)(1-\lambda^2)}\ .
\label{4.23}\end{equation}
When $z_1 \in (\lambda^\nu,\ 1)$ this expression is maximum at 
$z_1 = \lambda^{\nu/2}$, so that
\begin{equation}
\vhi'(1) \le {\lambda \over Z_1(\lambda)} = 
{\lambda \over (1+\lambda)(1+\sqrt{\lambda})^2} < {1\over 8}\ \ \ 
{\rm if}\ \ p \ge 2\ .
\label{4.23.1}\end{equation}
Therefore, if $p\ge2$ and we choose  
$b \ge b_0(r\nu) = (1/8)^{1/r\nu}$, then $\vhi'(1) < b^{r\nu}$.
For a slightly better choice of $b$, we note that 
$\lambda \mapsto \lambda/Z_1(\lambda)$ is increasing on $(0,\ 1)$,
so that 
$\lambda \le b \Rightarrow \vhi'(1) \le b/Z_1(b)$. This will be less 
than $b^{r\nu}$ if $b \ge b_1(r\nu)$, where 
$s \mapsto b_1(s)$ is the solution of $b_1^s = b_1/Z_1(b_1)$, i.e.
the inverse function (defined on $(1,\ \infty)$) of 
\begin{equation}
b \mapsto 1+ {\log Z_1(b) \over \log(1/b)} =
1 + {\log((1+b)(1+\sqrt{b})^2) \over \log(1/b)}\ .
\label{4.23.2}\end{equation}
This last function is strictly increasing on 
$(0,\ 1)$, and tends to 1 as $b$ tends to 0 and to $+\infty$ as
$b$ tends to 1. Obviously $b_1(s) \le b_0(s)$.
A useful inequality (proved in Appendix 1) is:
\begin{equation}
{\log((1+b)(1+\sqrt{b})^2) \over \log(1/b)}\ \ >\ \
{2b \over 1-b}\ \ \ \forall b \in (0,\ 1)
\label{4.23.2.1}\end{equation}
i.e.
\begin{equation}
b = b_1(s),\ \ s>1\ \ \ \Rightarrow 
s > {1+b \over 1-b}\ .
\label{4.23.2.2}\end{equation}
If $p=1$,
\begin{equation}
\vhi'(1) \le 
{\lambda (1-\lambda^\nu) \over (1-\lambda^2)} \le
{\lambda \over (1+\lambda)} \le
{1 \over 2},\ \ \ \ (p=1).
\label{4.23.3}\end{equation}
Thus, if $b$ is chosen at least equal to $b_2(r\nu) = (1/2)^{1/r\nu}$ or to 
$b_3(r\nu)$, where $b_3$ is the inverse function of 
$b \mapsto 1+\log(1+b)/\log(1/b)$, it follows from $\lambda \le b$ that
$\vhi'(1) \le b^{r\nu}$.

We now define the action of the operator $B(b,\ r,\ p,\ \nu)$ on $\Phi_0$
by
\begin{equation}
B(b,\ r,\ p,\ \nu)\,\Phi_0 = \Phi = 
h_{b,\lambda}^{-1} \circ \vhi \circ h_{b,\lambda}\ .
\label{4.23.4}\end{equation}
This definition implies that if $\Phi_0$ is a fixed point of
$B(b,\ r,\ p,\ \nu)$, i.e. $\Phi = \Phi_0$, then the functions
$\vhi_0$ and $\vhi$ constructed above coincide, and $\lambda$, $z_1$,
$\psi$, and $u(z) = z_1\lambda^{1-\nu}\psi(z)^{1/r}$ provide a solution 
to the problem set in Section \ref{state}. Conversely, given a solution to
the problem, the function $\Phi$ given by Eq. (\ref{4.23.4}) (with
$\lambda = \Phi'(1)^{1/r\nu}$) is a fixed point of $B(b,\ r,\ p,\ \nu)$
for any $b \in [\lambda,\ 1)$.

The preceding estimates show that if $p \ge 2$ and $b \ge b_0(r\nu)$
or $b \ge b_1(r\nu)$, or if $p=1$ and $b \ge b_2(r\nu)$ or $b \ge b_3(r\nu)$,
then
\begin{equation}
B(b,\ r,\ p,\ \nu)\,\QQ_0(b,\ r\nu)  \subset 
\QQ_0(b,\ r\nu)\ .
\label{4.23.5}\end{equation}
The same estimates show that, for {\it any} solution of our problem,
the inequalities $\lambda < b_j(r\nu)$ must hold ($j= 0,\ 1$ for
$p \ge 2$, $j= 2,\ 3$ for $p=1$).

The set $\QQ_0(b,\ r\nu)$ is not compact: we have to
guard against $\lambda$ tending to zero, i.e. to find a reproducing lower
bound for $\lambda$. This will be feasible only under certain 
restrictions on $r$, $\nu$, and $p$. We first show that such restrictions
are unavoidable. $\vhi'(1)$ is given by
\begin{eqnarray}
\vhi'(1) = \lambda^\nu\prod_{j=0}^{p-1} z_1 \lambda^{1-\nu} v'(\zeta_j) &=&
\lambda^\nu\prod_{j=0}^{p-1} 
{\zeta_{j+1} v'(\zeta_j) \over v(\zeta_j)} \nonumber \\
= \prod_{j=0}^{p-1} {\zeta_j v'(\zeta_j) \over v(\zeta_j)} &=&
\prod_{j=0}^{p-1} 
{-\zeta_j \psi'(-\zeta_j) \over r\,\psi(-\zeta_j)}\ .
\label{4.23.6}\end{eqnarray}
Here we have used $\zeta_p/\zeta_0 = \lambda^{-\nu}$. The upper bound in 
(\ref{2.6}) (with $u_- = \lambda$) give
\begin{eqnarray}
\vhi'(1) \le \prod_{j=0}^{p-1}
{\zeta_j (1+\lambda) \over
r (1+\zeta_j) (1-\lambda\zeta_j)} &\le&
{\lambda \over r (1-\lambda^2)}
\left({\lambda^{1-\nu}(1+\lambda) \over 
r (1+\lambda^{1-\nu})(1-\lambda^{2-\nu})}\right)^{p-1} \nonumber \\
&\le& \lambda^{1+(p-1)(1-\nu)}\ 
(r\,(1-\lambda))^{-p}\ .
\label{4.23.7}\end{eqnarray}
If we suppose $p \ge 2$ and $\lambda \le b_1(r\nu)$, then
$r(1-\lambda) \ge 1+\lambda > 1$ by the inequality (\ref{4.23.2.2}).
Therefore a fixed point can exist only if 
\begin{equation}
r\nu - 1-(p-1)(1-\nu) > 0\ .
\label{4.23.8}\end{equation}

Using (\ref{4.23.6}) and the lower bound in (\ref{2.6}), with 
$u_+ = a_0(\lambda) < b$ gives
\begin{equation}
\vhi'(1) \ge \prod_{j=0}^{p-1}
{\zeta_j (1-b) \over
r (1+\zeta_j) (1+b\zeta_j)}\ ,
\label{4.23.9}\end{equation}
and using $1 \ge \zeta_j \ge z_1 \lambda^{1-\nu}$ (for $j >0$), and
$z_1 \ge (1-b)/2$ (see (\ref{4.17})), we find
\begin{equation}
\vhi'(1) \ge c^p \lambda^{1+(p-1)(1-\nu)},\ \ \ \ 
c = {(1-b)^2 \over 4 r (1+b)}\ .
\label{4.23.10}\end{equation}
Assume now that $\lambda \ge \lambda_0 >0$. Then a sufficient condition
for $\vhi'(1) \ge \lambda_0^{r\nu}$ to hold is that
\begin{equation}
\lambda_0^{r\nu - 1 - (p-1)(1-\nu)} \le c^p \ .
\label{4.23.11}\end{equation}
If the condition (\ref{4.23.8}) 
holds, we can take
\begin{equation}
\lambda_0 = \lambda_0(p,\ r,\ \nu)  = 
\left ({(1-b)^2 \over 4 r (1+b)}\right)^{p \over r\nu - 1 - (p-1)(1-\nu)}
\ \ \ \in (0,\ 1).
\label{4.23.13}\end{equation}

Assume that the inequality (\ref{4.23.8}) holds.
Let, for definiteness, $b(r\nu) = b_1(r\nu)$ if $p \ge 2$, and
$b(r\nu) = b_3(r\nu)$ if $p =1$. We observe that 
\begin{equation}
\QQ_1(p,\ r,\ \nu) =
\QQ_0(b(r\nu),\ r\nu) \cap \{\Phi\ :\ \Phi'(1) \ge 
\lambda_0(p,\ r,\ \nu)^{r\nu}\}
\label{4.23.14}\end{equation}
is not empty. Indeed the function 
$\chi_{b,s}$ (see (\ref{2.2.2})) with $b = b(r\nu)$ and $s=r\nu$ belongs to
$\QQ_0(b(r\nu),\ r\nu)$ and $\chi'_{b,s} = b^{r\nu}$. Therefore
$\Phi = B(b(r\nu),\ r,\ p,\ \nu)\,\chi_{b,s}$ also belongs to 
$\QQ_0(b(r\nu),\ r\nu)$, and the preceding estimates show that 
$b^{r\nu} \ge \Phi'(1) \ge b^{1+(p-1)(1-\nu)} c^p$ with $c$ as in
(\ref{4.23.10}), and $b = b(r\nu)$. Hence $b(r\nu) \ge \lambda_0(p,\ r,\ \nu)$,
in particular $\chi_{b,s} \in \QQ_1(p,\ r,\ \nu)$.
(This is not really essential since we could have redefined 
$\lambda_0(p,\ r,\ \nu)$ to be less than $b(r\nu)$.)
The continuous map $B(b(r\nu),\ r,\ p,\ \nu)$ maps
the compact convex non-empty set $\QQ_1(p,\ r,\ \nu)$ into 
itself. Therefore it has a fixed point there by the 
Schauder-Tikhonov theorem. As noted before,
if $\Phi_0 = \Phi$ is such a fixed point, the functions $\vhi_0$ and
$\vhi$ constructed as above coincide, and $\psi$ and 
$u(z) = z_1\lambda^{1-\nu}\psi(z)^{1/r}$ provide a solution to our
problem.
Note that here again {\it any} solution 
must satisfy $\lambda \ge \lambda_0(p,\ r,\ \nu)$,
since it must satisfy 
$\vhi'(1) = \lambda^{r\nu} \ge c^p \lambda^{1+(p-1)(1-\nu)}$,
with $c$ as in (\ref{4.23.10}). Thus any solution is associated to
a fixed point of $B(b(r\nu),\ r,\ p,\ \nu)$ in $\QQ_1(p,\ r,\ \nu)$.

\section{Case $p=1$ and $\nu \in [1,\ 2]$}

\label{p=1nu=2}
This case has been dealt with in [E3]. It will be shown in this section
that the method of the preceding section also applies to this case with
minor modifications. Let $r > 1$ and $\nu \in [1,\ 2]$ be fixed reals.
We define the space $\QQ(b,\ r\nu)$ and the operator $B(b,\ r,\ 1,\ \nu)$ 
in the same way as in the preceding section. In particular,
starting from $\Phi_0 \in \QQ(b,\ r\nu)$, the functions
$\vhi_0$, $\psi$ and $v$ are defined by the same formulae and have the same
properties, in particular
\begin{equation}
v(-1) = 0,\ \ \ v(0) = 1,\ \ \ 
v(1/\lambda) \le \lambda^{-\nu}\ ,
\label{5.1}\end{equation}
but we note that now $\lambda^{-\nu} \ge 1/\lambda$. We define 
\begin{equation}
z_1 = 1/v(\lambda) \in (0,\ 1).
\label{5.2}\end{equation}
It follows from (\ref{5.1}) that 
\begin{equation}
z_1 > \lambda^\nu
\label{5.3}\end{equation}
and from the upper bound (\ref{2.5}) on $\psi(-\lambda)$ that
\begin{equation}
z_1 \ge (1-\lambda)^{1/r} > 1-\lambda\ .
\label{5.4}\end{equation}
The function 
\begin{equation}
z \mapsto \vhi(z) = z_1 v(\lambda z)
\label{5.5}\end{equation}
is again Herglotzian, holomorphic in $\Omega(-1/\lambda,\ 1/\lambda^2)$,
continuous on $[-1/\lambda,\ 1/\lambda^2]$ with
\begin{equation}
\vhi(-1/\lambda)\  = 0,\ \ \ 
\vhi(1) =1,\ \ \ 
\vhi(1/\lambda^2) = z_1 v(1/\lambda) \le z_1 /\lambda^\nu < 1/\lambda^2\ .
\label{5.6}\end{equation}
Schwarz's lemma can be again applied as in the preceding section, but now with
\begin{equation}
A = 1+{1\over\lambda},\ \ B = B' = {1\over\lambda^2} - 1,\ \ 
A' = 1\ .
\label{A.3}\end{equation}
This gives
\begin{equation}
\vhi'(1) \le \lambda\ ,
\label{A.4}\end{equation}
which is not sufficient for our purposes. 
We therefore use the bound
(\ref{2.7}) with $u_- = \lambda$ and $M = 1/\lambda^{r\nu}$, to get
\begin{equation}
{z v'(z) \over v(z)} = 
-{z \psi'(-z)\over r\,\psi(-z)} \le 
{\nu \log(1/\lambda) \over 4(1-\lambda z)}\ \ \ 
\forall z \in (0,\ 1/\lambda)\ ,
\label{A.5}\end{equation}
hence 
\begin{equation}
\vhi'(1)  = {\lambda v'(\lambda) \over v(\lambda)} \le  
{\log(1/\lambda) \over 2(1-\lambda^2)}\ .
\label{A.6}\end{equation}
The r.h.s. of this inequality is a decreasing function of $\lambda$, 
tending to $+\infty$ when $\lambda \rightarrow 0$, and to 
$1/4$ when $\lambda \rightarrow 1$.
For any choice of $b \in (0,\ 1)$, if $\lambda < b^{r\nu}$, then
$\vhi'(1) < b^{r\nu}$ by (\ref{A.4}). If $\lambda \ge b^{r\nu}$, then, by 
(\ref{A.6}), a sufficient condition for $\vhi'(1) < b^{r\nu}$ is that 
$b^{r\nu} > \mu$, where $\mu$ is  the unique zero, in 
$(0,\ 1)$, of the increasing function 
\begin{equation}
x \mapsto x - {\log(1/x)  \over 2(1-x^2)} \ .
\label{A.7}\end{equation}
One finds $\mu <0.479$, and we choose, from now on, 
$b = b_5(r\nu) = (0.479)^{1/r\nu}$. Note that for {\it any} solution, 
$\vhi'(1) = \lambda^{r\nu}$ must satisfy (\ref{A.6}), and since the rhs
of this inequality is decreasing, the function defined in (\ref{A.7})
must be negative at $x=\lambda^{r\nu}$, i.e. 
$\lambda \le b_5(r\nu)$.

The lower bound in (\ref{2.6}), with $u_+ = a_0(\lambda) <b$, gives
\begin{equation}
\vhi'(1) = -{\lambda \psi'(-\lambda) \over r\psi(-\lambda)} \ge
{\lambda (1-a_0(\lambda))\over r\,(1+\lambda)(1+\lambda a_0(\lambda))}
\ge {\lambda (1-b) \over r\,(1+b)(1+b^2)}\ .
\label{A.8}\end{equation}
If $\lambda \ge \lambda_0 > 0$ then $\vhi'(1) \ge \lambda_0^{r\nu}$
provided $\lambda_0 \le \lambda_0(r,\ \nu)$ with
\begin{equation}
\lambda_0(r,\ \nu) = 
\left ({(1-b) \over r\,(1+b)(1+b^2)} \right)^{1/(r\nu -1)}\ ,
\ \ \ \ \ \ b = b_5(r\nu)\ .
\label{A.9}\end{equation}
Therefore the operator $B(b_5(r\nu),\ r,\ 1,\ \nu)$ 
preserves the compact convex set
\begin{equation}
\QQ_0(b_5(r\nu),\ r\nu) \cap 
\{\Phi\ :\ \Phi'(1) \ge \lambda_0(r,\ \nu)\}.
\label{A.10}\end{equation}
Again {\it any} solution must satisfy $\lambda \ge \lambda_0(r,\ \nu)$.

\section {Case $p\ge 2$ and $\nu \in [1,\ 2]$}

\label{p>1nu>1}
In this section, $r > 1$ and $\nu \in [1,\ 2]$ are fixed real numbers,
and $p \ge 2$ is a fixed integer. 
The real number $b \in [1/2,\ 1)$ is also fixed, but its value
will be chosen later (as a function of $r$). We shall
need the function $a: [0,\ 1] \rightarrow [0,\ 1]$ given by
\begin{equation}
a(t) = \min \left \{ {2t \over 1-t}\ ,\ {1+t \over 2} \right \}
= \ \left \{ \matrix{
{\displaystyle 2t \over \displaystyle \strut 1-t}
&\ \ {\rm if}\ \ & 0 \le t \le \sqrt{5}-2\ ,\cr
{\displaystyle 1+t \over \displaystyle \strut 2}
&\ \ {\rm if}\ \ & \sqrt{5}-2 \le t \le 1\ .\cr}
\right. 
\label{6.1}\end{equation}
$t\mapsto a(t)$ is continuous and strictly increasing in $[0,\ 1]$. 
(Note that $\sqrt{5}-2 \approx 0.236 < 1/4$.)

We denote $\wt{\QQ}_0(b,\ \nu r)$ the space of all functions $\Phi$ with the
following properties:

\begin{description}
\item{(\~Q1)}\quad  $\Phi$ is a Pick function holomorphic in the domain:
\begin{equation}
\Omega(-1/b,\ 1/a(b)) = \bC_+ \cup \bC_- \cup
\left (-{1\over b},\ {1\over a(b)} \right )\ ,\ \ \ 
a(b) = {1+b \over 2}\ ,
\label{6.2}\end{equation}
and maps this domain into itself.

\item{(\~Q2)}\quad  $\Phi(z) \ge 0$ for all $z \in (-1/b,\ 1/a(b))$,

\item{(\~Q3)}\quad  $\Phi(1) = 1$, and $0< \Phi'(1) \le b^{\nu r}$. 
\end{description} 

\noindent $\wt{\QQ}_0(b,\ \nu r)$ is a convex subset of the real Fr\'echet
space of all self-conjugated functions holomorphic in $\Omega(-1/b,\ 1/a(b))$.
It is not empty since it contains
the function $\chi_{b,\nu r}$ (see Section \ref{prel}).

\subsection{The operator $B(b,\ r,\ p,\ \nu)$}

\label{opB}
We shall define a continuous
operator $B(b,\ r,\ p,\ \nu)$ on the space $\wt{\QQ}_0(b,\ \nu r)$
by describing its action on an
arbitrary element $\Phi_0$.

Given $\Phi_0 \in \wt{\QQ}_0(b,\ \nu r)$, we denote 
$\lambda = \Phi_0'(1)^{1/r\nu}$.  Note that $\lambda \le b$.
We define a function $\vhi_0$ by
\begin{equation}
\vhi_0 = h_{b,\lambda}\circ\Phi_0\circ h_{b,\lambda}^{-1}\ .
\label{6.4}\end{equation}
Here $h_{b,\lambda}$ is the homographic function defined 
in Section \ref{crpnu} (see (\ref{4.1})). It maps the domain
$\Omega(-1/b,\ 1/a(b))$ onto
$\Omega(-1/\lambda,\ 1/a_1(\lambda))$, where
$1/a_1(\lambda) = h_{b,\lambda}(1/a(b))$, i.e.
\begin{equation}
a_1(\lambda) = {1+\lambda \over 2} + {b-\lambda \over 1+b}\ \ 
\ge a(b)\ \ \ge \ \ a(\lambda),\ \ \ \ \ 
a_1(\lambda) \ \le \ a_1(0) = {1+3b \over 2(1+b)}\ .
\label{6.5}\end{equation}
The function $\vhi_0$ possesses the following properties:

\begin{description}
\item{(\~Q$'1$)}\quad  $\vhi_0$ is a Pick function holomorphic in
$\Omega(-1/\lambda, 1/a_1(\lambda))$, and maps this domain into itself.

\item{(\~Q$'2$)}\quad  $\vhi_0(z) \ge 0$ for all 
$z \in (-1/\lambda,\ 1/a_1(\lambda))$.

\item{(\~Q$'3$)}\quad  $\vhi_0(1) = 1$, and $\vhi'_0(1) = \lambda^{\nu r}$.
\end{description}

\noindent
As in previous sections  
we denote $\psi$ the linearizer of $\vhi_0$, normalized by the 
condition $\psi(0) = 1$, i.e. 
\begin{equation}
\psi(z) = {1\over \lambda^{\nu r}} \psi(\vhi_0(z)) \ \ \ 
\forall z \in \Omega(-1/\lambda, 1/a_1(\lambda)),\ \ \ 
\psi(1) = 0,\ \ \ \psi(0) = 1\ .
\label{6.6}\end{equation}
$\psi$  is anti-Herglotz, holomorphic in $\Omega(-1/\lambda, 1/a_1(\lambda))$, 
and satisfies the inequalities (\ref{2.5}) and (\ref{2.6}), 
with $u_- = \lambda$ and $u_+ = a_1(\lambda)$. Also,
\begin{equation}
\psi(-1/\lambda) = {1 \over \lambda^{\nu r}} \psi(\vhi_0(-1/\lambda))\ \ 
\le {1 \over \lambda^{\nu r}}\ .
\label{6.7}\end{equation}
We again define $v(z) = (\psi(-z))^{1/r}$ fo all 
$z\in \Omega(-1,\ 1/\lambda)$. This is a Pick function which 
extends to a strictly increasing continuous function on 
$[-1,\ 1/\lambda]$. It satisfies:
\begin{equation}
v(-1) = 0,\ \ \ v(0) = 1,\ \ \ 
v(1/\lambda) \le \lambda^{-\nu}\ .
\label{6.9}\end{equation}
We now show that there is a unique $z_1 \in (0,\ 1)$ such that
\begin{equation}
\left ( {z_1 \over \lambda^{\nu-1}}v \right )^p (\lambda) = 
{1 \over \lambda^{\nu-1}}\ .
\label{6.10}\end{equation}
For real $s \ge 0$ let $x_0(s) = \lambda$, 
$x_1(s) = s\lambda^{1-\nu}v(\lambda)$. The function $s \mapsto x_1(s)$
is strictly increasing on $\bR_+$ and takes the values $\lambda$ at
$s_* = \lambda^\nu/v(\lambda)$ and $1/\lambda$ at
$s_1 = \lambda^{\nu-2}/v(\lambda)$. By induction we can construct
a strictly decreasing infinite sequence 
$s_1 > \dots > s_j > \dots > s_*$ such that, for $j \ge 2$, 
$s\mapsto x_j(s) = (s\lambda^{1-\nu}v)^j(\lambda)$ is continuous and
strictly increasing on $[s_*,\ s_{j-1}]$, 
$x_0(s) < \dots < x_j(s)$ in $(s_*,\ s_{j-1}]$, 
$x_j(s_*) = \lambda$, and $x_j(s_j) = 1/\lambda$. 
Indeed it follows that $x_{j+1}(s) = s\lambda^{1-\nu}v(x_j(s))$
is defined, continuous, and strictly increasing on
$[s_*,\ s_j]$ and 
$x_{j+1}(s) > s\lambda^{1-\nu}v(x_{j-1}(s)) = x_j(s)$ for all
$s \in (s_*,\ s_j]$.
Since $x_{j+1}(s_j) > x_j(s_j) = 1/\lambda$ and 
$x_{j+1}(s_*) = \lambda$, $s_{j+1}$ exists in $(s_*,\ s_j)$.
In particular $x_p(s_{p-1}) > 1/\lambda$. 
Therefore there is a unique $z_1 \in (s_*,\ s_{p-1})$ such that
$x_p(z_1) = \lambda^{1-\nu}$. It must satisfy $z_1 <1$ since
$z_1v(x_{p-1}(z_1)) = 1$, and $v(x_{p-1}(z_1)) >1$.
Note also that for $s \in (s_*,\ s_{p-1})$, there exists a unique
$x_{-1}(s) < x_0(s) = \lambda$ such that 
$s\lambda^{1-\nu}v(x_{-1}(s)) = \lambda$. 
The function 
$z \mapsto s_*\lambda^{1-\nu} v(z)$ maps $\Omega(-1,\ 1/\lambda)$
into $\Omega(0,\ \lambda^{1-\nu}/v(\lambda))$, so that it
has a  unique and attractive fixed point at 
$\lambda$ by Schwarz's lemma. 
Hence $s_*\lambda^{1-\nu} v(x) \ge x$ 
for all $x \in [-1,\ \lambda]$. When $s > s_*$, 
$s\lambda^{1-\nu}v(x) > x$ 
for all $x \in [-1,\ \lambda]$. Since this includes
$[x_{-1}(s),\ x_0(s)]$, it follows that 
$s\lambda^{1-\nu}v(x) > x$ for all $x \in [-1, x_p(s)]$, 
for all $s \in (s_*,\ z_1]$. The function $x_p$ is analytic,
with a strictly positive derivative, on $(s_*,\ s_{p-1})$.
Therefore $z_1$ depends continuously on $v$, hence on $\vhi_0$.

We denote $\zeta_j = (z_1 \lambda^{1-\nu} v)^j (\lambda)$, 
$(0 \le j \le p+1)$ :
\begin{equation}
\lambda = \zeta_0 < \zeta_1 < \dots \zeta_p = {1 \over \lambda^{\nu-1}} < 
\zeta_{p+1} = {z_1 \over \lambda^{\nu-1}} v(1/\lambda^{\nu-1}),
\label{6.11}\end{equation}
Since $v(1/\lambda^{\nu-1}) \le \lambda^{-\nu}$, 
\begin{equation}
z_1  > \lambda^\nu\ ,\ \ \ \ {z_1 \over \lambda^{\nu-1}} > \lambda\ ,
\label{6.12}\end{equation}
and since $v(\lambda) > v(0) = 1$,
\begin{equation}
\zeta_1 > {z_1 \over \lambda^{\nu-1}}\ .
\label{6.13}\end{equation} 
We have seen above that
\begin{equation}
{z_1 \over \lambda^{\nu-1}} v(x) > x \ \ \
\forall x \in [-1,\ 1/\lambda^{\nu-1}]\ .
\label{6.14}\end{equation}
Applying this to $x = 1$ gives $z_1/\lambda^{\nu-1} \ge 1/v(1)$, and, using
(\ref{2.5}),
\begin{equation}
{z_1 \over \lambda^{\nu-1}} \ \ \ge \ \ 
\left ( {1 - \lambda \over 2} \right )^{1\over r}\ \ > \ \ 
{1 - \lambda \over 2}\ .
\label{6.15}\end{equation}
The function $\vhi$ is defined by:
\begin{equation}
\vhi(z) = \lambda^{\nu-1}
\left ( {z_1 \over \lambda^{\nu-1}} v \right )^p (\lambda z)\ ,
\ \ \ z\in \bC_+ \cup \bC_- \cup (-1/\lambda,\ \zeta_1/\lambda)\ .
\label{6.16}\end{equation}
In this domain, $\vhi$ is a Pick function, which extends continuously
to the ends of its real interval of definition, and
\begin{eqnarray}
\vhi(-1/\lambda)\  &=&\  
\lambda^{\nu-1} \left ( {z_1 \over \lambda^{\nu-1}} v \right )^{p-1}(0)
\ \ge \ z_1\ \ge\ \lambda^\nu\ \ge\ \lambda^2\ ,\nonumber \\
\vhi(1)\ &=&\ 1\ ,\ \ \ \ 
\vhi(\zeta_1/\lambda)\ =\ z_1 v(1/\lambda^{\nu-1})\ \le\
{z_1 \over \lambda^\nu}\ 
<\  \zeta_1/\lambda\ .
\label{6.17}\end{eqnarray}
(Note that the first inequality in (\ref{6.17}) has used $p \ge 2$.)
The domain $\Omega(-1/\lambda,\ \zeta_1/\lambda)$ is a basin of
attraction of the fixed point 1 of $\vhi$, hence
$\vhi'(1) < 1$ by Schwarz's lemma. For a better upper bound on this
derivative, we shall need a better lower bound for $\zeta_1/\lambda$.
This is provided by 

\begin{lemma}\label{clowb}
The inequality
\begin{equation}
{z_1 \over \lambda^{\nu-1}} v(z) \ge
{z(1-2\lambda) + \lambda \over 1 - \lambda z}
\label{6.18}\end{equation}
holds for all $z \in [0,\ 1]$.
\end{lemma}

\PROOF
This is simply the result of 
applying Lemma \ref{noses} of Section \ref{prel}, 
with $f=z_1\lambda^{1-\nu}v$, and 
$a=0$, $b=1$, $B = 1/\lambda$. This function satisfies 
$f(0) = z_1\lambda^{1-\nu} \ge \lambda$
by (\ref{6.12}), and $f(1) \ge 1$ by (\ref{6.14}).

For $z = \zeta_0 = \lambda$, this
gives $\zeta_1 \ge 2\lambda/(1+\lambda)$. Since we also have the lower bounds
(\ref{6.13}) and (\ref{6.15}),
\begin{equation}
{\zeta_1 \over \lambda} \ge
\max \left\{{2 \over 1+\lambda},\ {1-\lambda \over 2\lambda}\right \}\ \
=\ \ {1\over a(\lambda)}\ . 
\label{6.19}\end{equation}
This is the reason for our original definition of the function $a$ in
(\ref{6.1}). We conclude that the domain 
$\Omega(-1/\lambda,\ \zeta_1/\lambda)$ where $\vhi$ is holomorphic,
and which it maps into itself, certainly contains the domain of 
analyticity $\Omega(-1/\lambda,\ 1/a_1(\lambda))$ of $\vhi_0$ in view of
(\ref{6.5}).
We now use Schwarz's lemma, as mentioned in Section \ref{prel}, to
obtain an upper bound for $\vhi'(1)$ :
\begin{eqnarray}
\vhi'(1) \le {A'B'(A+B)\over AB(A'+B')}\ \ &{\rm with}&\nonumber \\
A = 1+{1\over\lambda}&,& B = B' = {1\over a(\lambda)} - 1,\ \ \ 
A' = 1-\lambda^2\ .
\label{6.20}\end{eqnarray}
This gives
\begin{eqnarray}
\vhi'(1) &\le& 
{(1-\lambda^2)(a(\lambda)+\lambda) \over (1+\lambda)(1-a(\lambda)\lambda^2)}
\nonumber \\
&\le&\ \ Z(\lambda) = {1+3\lambda \over 2+2\lambda +\lambda^2}\nonumber \\
&\le& Z_{\rm max} = 9/(4+2\sqrt{13}) < 0.803.
\label{6.21}\end{eqnarray}
Therefore choosing
$b \ge b_6(r\nu) = m_0^{1/r\nu}$, $m_0 = 0.803\,$ ensures that
$\vhi'(1) < b^{r\nu}$.

We define the operator $B(b,\ r,\ p,\ \nu)$ by 
\begin{equation}
\Phi = B(b,\ r,\ p,\ \nu) \Phi_0 
= h_{b,\lambda}^{-1} \circ \vhi \circ h_{b,\lambda}\ .
\label{6.22}\end{equation}
It then follows from the preceding estimates that, if $b \ge b_6(r\nu)$,
\begin{equation}
B(b,\ r,\ p,\ \nu) \,\wt{\QQ}_0(b,\ \nu r)  \subset 
\wt{\QQ}_0(b,\ \nu r)\ .
\label{6.23}\end{equation}
In the remainder of this section, it will always be understood
that $b = b_6(r\nu)$.

\subsubsection{Lower bound for $\vhi'(1)$.}

\noindent
We use
\begin{equation}
\vhi'(1) = \prod_{j=0}^{p-1} 
{-\zeta_j \psi'(-\zeta_j) \over r\,\psi(-\zeta_j)}\ .
\label{6.24}\end{equation}
The lower bound in (\ref{2.6}) gives, for $\zeta \in [0,\ 1/\lambda)$,
\begin{equation}
{-\zeta \psi'(-\zeta) \over \psi(-\zeta)} \ge 
{\zeta(1-c) \over (1+\zeta) (1+c\zeta)}\ ,
\label{6.25}\end{equation}
Here $c = a_1(\lambda)$, where $a_1(\lambda)$ is given by (\ref{6.5})
and satisfies
\begin{equation}
a(\lambda) \le a(b) \le a_1(\lambda) < a_1(0) = {1+3b \over 2(1+b)}\ .
\label{6.26}\end{equation}
However it will be convenient to suppose only, at first,
that (\ref{6.25}) holds for a certain $c$ satisfying
$a(\lambda) \le c <1$.
For $j > 0$, $\zeta_j \ge \zeta_1 \ge \lambda/a(\lambda)$ hence 
$\zeta_j \in [\lambda/c,\ 1/\lambda]$. When $\zeta$
varies in this interval, the second expression in (\ref{6.25}) is 
minimum at $\zeta = 1/\lambda$. Therefore
\begin{equation}
\vhi'(1) \ge \lambda^p \,
\left (
{1-c \over r\,(1+\lambda)(1+c\lambda)}\right )
\left(
{1-c \over r\,(1+\lambda)(\lambda +c)}\right )^{p-1}\ .
\label{6.27}\end{equation}
It is easy to verify that the rhs of this inequality is decreasing in $c$ 
and increasing in $\lambda$ provided $c \ge \lambda^2$ (note that
$a(\lambda) > \lambda$).
Setting now $c = a_1(\lambda)$, and
using the inequalities (\ref{6.26}) and $\lambda \le b$, this gives
\begin{equation}
\vhi'(1) \ge \lambda^p \,\left( {1-b \over 16r}\right )^p\ .
\label{6.28}\end{equation}
Supposing $\lambda \ge \lambda_0 > 0$, the last inequality will imply 
$\vhi'(1) \ge \lambda_0^{r\nu}$ 
if $\lambda_0$ satisfies 
\begin{equation}
\lambda_0^{r\nu -p} \le 
\left ( {1-b \over 16r}\right )^p\ ,
\label{6.28a}\end{equation}
and this is 
possible only if $r\nu -p > 0$. In this case we 
can choose 
\begin{equation}
\lambda_0 = \lambda_0(r,\ \nu) =
\left ( {1-b \over 16r}\right )^{p \over r\nu-p}\ ,
\label{6.29}\end{equation}
and obtain the existence of a
fixed point in the same way as in the preceding sections.
Recall that in these formulae, $b$ stands for
$b_6(r\nu) = m_0^{1/r\nu}$. It is easy to verify that
$\lambda_0(r,\ \nu) \rightarrow 1$ when $r \rightarrow \infty$.

The condition $r\nu -p > 0$ is just a limitation of the 
present method. 
The inadequacy of the estimate (\ref{6.28}) is due to
the fact that $a_1(\lambda)$ does not tend to 0 as $\lambda$ tends to
0. By contrast, in the case
of fixed points, the lower bound on $\lambda$ can be improved. Indeed,
since $\psi$ and $\vhi$ are holomorphic in 
$\Omega(-1/\lambda,\ 1/a(\lambda))$, the bound (\ref{6.25}) and 
consequently (\ref{6.27}) hold with $c$ replaced by $a(\lambda)$
(instead of $a_1(\lambda)$).
Assume $\lambda \le 1/7$. We can then set
$c = 2\lambda/(1-\lambda)$ in (\ref{6.27}) and obtain
\begin{equation}
\vhi'(1) \ge \lambda \,
\left (
{1-3\lambda \over r\,(1+\lambda)(1-\lambda+2\lambda^2)}\right )
\left(
{1-3\lambda \over r\,(1+\lambda)(3-\lambda)}\right )^{p-1}\ >\ 
\lambda\,(6r)^{-p}\ .
\label{6.31}\end{equation}
Therefore the lower bound
\begin{equation}
\lambda\ =\ \vhi'(1)^{1/r\nu} \ge \min\{1/7,\ (6r)^{-p/(r\nu - 1)}\}
\label{6.32}\end{equation}
holds for all fixed points. 

This fact suggests the use of another operator instead of
$B(b,\ r,\ p,\ \nu)$, and this will be done in the next subsection.

\subsection{The operator $N(b,\ r,\ p,\ \nu,\ \lambda_1)$}

\label{opN}
In this subsection, we define a new operator $N(b,\ r,\ p,\ \nu,\ \lambda_1)$
on the space $\wt{\QQ}_0(b,\ \nu r)$. This construction closely
follows an idea of Mestel and Osbaldestin [MO]. It consists in
replacing the operator $B(b,\ r,\ p,\ \nu)$ (which is analytic on
$\wt{\QQ}_0(b,\ \nu r)$) by a ``truncated version'' 
$N(b,\ r,\ p,\ \nu,\ \lambda_1)$ which is only continuous, 
but maps $\wt{\QQ}_0(b,\ \nu r)$ into a compact subset.
This operator depends on an additional real parameter 
$\lambda_1 \in (0,\ 1/2)$. It will be shown later that
for small values of this parameter, any fixed point of
$N(b,\ r,\ p,\ \nu,\ \lambda_1)$ is a fixed point of 
$B(b,\ r,\ p,\ \nu)$.

The notations are the same as in the preceding subsection
unless explicitly mentioned. In particular $\nu \in [1,\ 2]$
and $r > 1$ are fixed and $b$ will stand for 
$b_6(r\nu) = m_0^{1/r\nu}$, $m_0 = 0.803$ . 
We denote $\tau_1 = \lambda_1^r$.

We define $N(b,\ r,\ p,\ \nu,\ \lambda_1)$ by its action on an arbitrary
element $\Phi_0$ of $\wt{\QQ}_0(b,\ \nu r)$. Let
$\sigma^\nu = \Phi_0'(1)$. Recall that, by the definition of
$\wt{\QQ}_0(b,\ \nu r)$, $\sigma^\nu  \le b^{r\nu}$.

If $\sigma^\nu  \ge \tau_1^\nu$, we define 
\begin{equation}
N(b,\ r,\ p,\ \nu,\ \lambda_1) \Phi_0 = B(b,\ r,\ p,\ \nu) \Phi_0,
\ \ \ \ (\sigma^\nu  \ge \tau_1^\nu).
\label{6b.1}\end{equation}

If $\sigma^\nu  < \tau_1^\nu$, we define $\lambda = \lambda_1$
(so that $\lambda \le b$), and define $\vhi_0$, as before, by
\begin{equation}
\vhi_0 = h_{b,\lambda}\circ\Phi_0\circ h_{b,\lambda}^{-1}\ .
\label{6b.2}\end{equation}
The function $\vhi_0$ is holomorphic and Herglotzian in the domain
$\Omega(-1/\lambda,\ 1/a_1(\lambda))$, which it maps into itself. Here
$a_1(\lambda) = a_1(\lambda_1)$ is given by (\ref{6.5}).
$\vhi_0$ possesses the same properties as in  Subsection \ref{opB}, except for
\begin{equation}
\vhi_0'(1) = \sigma^{\nu}\ .
\label{6b.3}\end{equation}
The linearizer $\psi_1$ is the unique function holomorphic
in $\Omega(-1/\lambda,\ 1/a_1(\lambda))$ such that
\begin{equation}
\psi_1(z) = {1\over \sigma^{\nu}} \psi_1(\vhi_0(z))
\ \ \ \ \forall z \in \Omega(-1/\lambda,\ 1/a_1(\lambda)),
\ \ \ \ \psi_1(0) = 1,\ \ \psi_1(1) = 0\ .
\label{6b.4}\end{equation}
It is anti-Herglotzian and satisfies 
\begin{equation}
\psi_1(-1/\lambda) = {1 \over \sigma^{\nu}} \psi_1(\vhi_0(-1/\lambda))\ \ 
\le {1 \over \sigma^{\nu}}\ .
\label{6.b5}\end{equation}
In the preceding subsection, much depended on the bound 
$\psi(-1/\lambda) \le \lambda^{-r\nu}$.
To restore an analogous situation 
we define a new function $\psi$ as
\begin{equation}
\psi = \theta_{\sigma^{-\nu},\tau_1^{-\nu}} \circ \psi_1\ ,
\label{6.b6}\end{equation}
where $\theta_{\sigma^{-\nu},\tau_1^{-\nu}}$ denotes the homographic
function which fixes 0 and 1, and sends $\sigma^{-\nu}$ to
$\tau_1^{-\nu}$:
\begin{equation}
\theta_{\sigma^{-\nu},\tau_1^{-\nu}}(z) =
{z(1-\sigma^\nu) \over z(\tau_1^\nu -\sigma^\nu) + 1-\tau_1^\nu} \ .
\label{6.b7}\end{equation}
This function is Herglotzian and has a pole at a negative value 
temporarily denoted $k$. As a consequence $\psi$ is holomorphic
and anti-Herglotzian in $\Omega(-1/\lambda,\ 1/a_2)$, where
$1/a_2 = \psi_1^{-1}(k)$ if $k \in \psi_1((1,\ 1/a_1(\lambda)))$,
and $1/a_2 = 1/a_1(\lambda)$ otherwise.
For $z >1$,
$\psi_1(z) <0$ and (using the inequalities (\ref{2.5})),
\begin{equation}
\psi_1(z) \ge \psi_2(z) = {1 - z \over 1- a_1(\lambda)z}
\label{6.b8}\end{equation}
If $y = \psi_1^{-1}(k) < 1/a_1$, we have, since $\psi_2$ is decreasing,
\begin{equation}
k = \psi_1(y) \ge \psi_2(y),\ \ \ \ \psi_2^{-1}(k) \le y\ .
\label{6.b8a}\end{equation}
Thus $\psi$ is holomorphic in
$\Omega(-1/\lambda,\ 1/\ell)$, where 
$1/\ell = \psi_2^{-1}(k) = \psi_2(k)$. This gives:
\begin{equation}
\ell = {\tau_1^\nu - \sigma^\nu + (1 - \tau_1^\nu) a_1(\lambda)
\over 1 -  \sigma^\nu},
\label{6.b9}\end{equation}
\begin{equation}
a(\lambda)\ \ \le\ \  a_1(\lambda)\ \ \le\ \ \ell\ \ <\ \ a_3(\lambda)
\ \ = \ \ \tau_1^\nu + (1 - \tau_1^\nu) a_1(\lambda)\ .
\label{6.b9.1}\end{equation}
The function $\psi$ has been defined so as to satisfy 
$\psi(-1/\lambda) \le \lambda ^{-r\nu}$.
We now proceed to define $v$, $z_1$, $\vhi$ etc. exactly as in
the preceding subsection and obtain the same inequalities with
the single exception that, in the lower bound (\ref{6.27}),
$c$ must be replaced by $a_3(\lambda)$. Since 
$\lambda = \lambda_1$, we find
\begin{equation}
\vhi'(1) \ge l(\lambda_1) = \lambda_1^p \,
\left (
{1-a_3(\lambda_1) \over r\,(1+\lambda_1)(1+a_3(\lambda_1)\lambda_1)}\right )
\left(
{1-a_3(\lambda_1) \over r\,(1+\lambda_1)
(\lambda_1 +a_3(\lambda_1))}\right )^{p-1}\ .
\label{6.b10}\end{equation}
Recall that $\vhi$ is holomorphic in 
$\Omega(-1/\lambda,\ 1/a(\lambda))$ and maps this domain into
itself, with $a(\lambda)$ given by (\ref{6.1}).
The bound (\ref{6.b10}) also holds in the cases when 
$\lambda > \lambda_1$ since then $a_1(\lambda) < a_3(\lambda_1)$.

Finally we define
\begin{equation}
N(b,\ r,\ p,\ \nu,\ \lambda_1)\Phi_0 =
h_{b,\lambda}^{-1} \circ \vhi \circ h_{b,\lambda}\ .
\label{6.b11}\end{equation}
The operator $N(b,\ r,\ p,\ \nu,\ \lambda_1)$ 
maps the domain $\wt{\QQ}_0(b,\ \nu r)$ into
$\wt{\QQ}_0(b,\ \nu r) \cap \{\Phi\ :\ \Phi'(1) \ge l(\lambda_1)\}$,
which is compact and convex, hence it has fixed points there.

Our task is now to prove that if $\lambda_1$ has been chosen
sufficiently small, any fixed point of $N(b,\ r,\ p,\ \nu,\ \lambda_1)$ is 
actually a fixed point of $B(b,\ r,\ p,\ \nu)$. We assume, from now on,
that $\lambda_1 \le 1/8$.
Let $\Phi_0$ be a fixed point of $N(b,\ r,\ p,\ \nu,\ \lambda_1)$.
If $\sigma^\nu = \Phi_0'(1) \ge \tau_1^\nu$, there is nothing
to prove. Otherwise, we have $\lambda = \lambda_1$ and
$\vhi_0 = \vhi$, so that $\vhi_0$ and
$\psi_1$ are now holomorphic in $\Omega(-1/\lambda_1,\ 1/a(\lambda_1))$.
Thus $\psi$ is now holomorphic in $\Omega(-1/\lambda_1,\ 1/a_4(\lambda_1))$,
with
\begin{equation}
a(\lambda_1)\ \ <\ \ 
a_4(\lambda_1) = {\tau_1^\nu - \sigma^\nu + (1 - \tau_1^\nu) a(\lambda_1)
\over 1 -  \sigma^\nu} 
\ \ < \ \ \tau_1^\nu + (1 - \tau_1^\nu) a(\lambda_1)\ .
\label{6.b12}\end{equation}
Recalling that $\lambda_1 \le 1/8$, we find
\begin{equation}
a_4(\lambda_1)\ \  \le\ \  \lambda_1^{r\nu} + 2\lambda_1{
1-\lambda_1^{r\nu} \over 1-\lambda_1}\ \  \le\ \ 
{3\lambda_1 \over 1-\lambda_1}
\label{6.b13}\end{equation}
Inserting this in the lower bound obtained by setting $\lambda = \lambda_1$
and $c = a_4(\lambda_1)$ in (\ref{6.27}) gives
\begin{equation}
\vhi'(1) \ge \lambda_1 \,
\left (
{1-4\lambda_1 \over r\,(1+\lambda_1)(1-\lambda_1+3\lambda_1^2)}\right )
\left(
{1-4\lambda_1 \over r\,(1+\lambda_1)(4-\lambda_1)}\right )^{p-1}\ ,
\label{6.b14}\end{equation}
and, using $\lambda_1 \le 1/8$,
\begin{equation}
\vhi'(1) \ge \lambda_1 (9r)^{-p},
\label{6.b14a}\end{equation}
and since $\lambda_1 \ge (\vhi'(1))^{1/r\nu}$,
\begin{equation}
\vhi'(1) \ge (9r)^{-pr\nu/(r\nu - 1)}\ .
\label{6.b15}\end{equation}
If we assume that $\lambda_1$ has been chosen so that
\begin{equation}
\lambda_1 < (9r)^{-p/(r\nu - 1)}\ ,
\label{6.b16}\end{equation}
the inequality (\ref{6.b15}) contradicts our hypothesis that
$\Phi_0'(1) < \lambda_1^{r\nu}$. Therefore $\Phi_0$ is a fixed point 
of $B(b,\ r,\ p,\ \nu)$. 

\section{Properties of solutions}

\label{Prop}
This section is devoted to some properties of
the solutions, i.e. of functions $\psi$ and $u$, and numbers 
$\nu \in (0,\ 1]$, $p \ge 2$, $r>1$, ($r\nu > 1+(p-1)(1-\nu)$ if $\nu < 1$), 
$\lambda$, $z_1$, satisfying the requirements of Section \ref{state}. 
These properties are extensions of those established for $p =1$ in 
\cite{EE,EL,E2}.
We do not consider the case $p =1$.

We denote 
$\vhi = \vhi_0$, $v$, $\zeta_0,\dots,\zeta_{p+1}$,  
the objects constructed from 
$\psi$ as in the definition of $B(b,\ r,\ p,\ \nu)$. We also 
denote $\tau = \lambda^r$, and
\begin{eqnarray*}
u(z) = \check u(-z) &=& 
{z_1 \over \lambda^{\nu-1}}\,v(-z) \\
&=&
{z_1 \over \lambda^{\nu-1}}\, \psi(z)^{1/r}\ \ = \ \ 
U(z)^{1/r}\ ,\ \ \ \ z \in \Omega(-1/\lambda,\ 1)\ ,
\end{eqnarray*}
\begin{equation}
U(z) = \ \ \left ({z_1 \over \lambda^{\nu-1}}\right )^r \psi(z)\ ,
\ \ \ \ z \in \Omega(-1/\lambda,\ \zeta_1/\lambda)\ .
\label{p.1}\end{equation}
Recall that it has been shown in Section \ref{crpnu} that 
\begin{equation}
{1 \over \tau^\nu} \ge {1\over \lambda}(1+\lambda)(1+\sqrt{\lambda})^2
> 8,\ \ \ \ 
\lambda \le b_1(r\nu),
\ \ \ \ \ {\rm if\ } 0<\nu \le 1,\ \ \ p \ge 2,
\label{p.2}\end{equation}
and 
\begin{equation}
r\nu \ge {1+\lambda \over 1-\lambda},\ \ \ \ \ 
\lambda^{r\nu -1 -(p-1)(1-\nu)} \le (1+\lambda)^{-p}
\ \ \ \ \ {\rm if\ } 0<\nu \le 1,\ \ \ p \ge 2.
\label{p.3}\end{equation}
Moreover (\ref{4.23.13}) and $r\nu \ge (1+b)/(1-b)$ give
\begin{equation}
\lambda \ge (4r^3\nu^2)^{-p/(r\nu -1 -(p-1)(1-\nu))}
\ \ \ \ \ {\rm if\ } 0<\nu \le 1,\ \ \ p \ge 2.
\label{p.3a}\end{equation}
For $1< \nu \le 2$, it was shown in Section \ref{p>1nu>1} that
\begin{equation}
\lambda \le b_6(r\nu) = m_0^{1/r\nu},\ \ \ m_0 = 0.803,\ \ \ \ \ \ 
{\zeta_1 \over \lambda} \ge {1\over a(\lambda)},\ \ \ \ 
(1 <\nu \le 2,\ \ \ p \ge 2)\ ,
\label{p.4}\end{equation}
where $a(\lambda)$ is defined in (\ref{6.1}), and that
\begin{equation}
\lambda\ \ge \min\{1/7,\ (6r)^{-p/(r\nu - 1)}\},\ \ \ \ 
(1 <\nu \le 2,\ \ \ p \ge 2)\ ,
\label{p.5}\end{equation}
\begin{equation}
\lambda\ \ge 
\left({1-m_0^{1/r\nu} \over 16r}\right)^{p/(r\nu -p)},\ \ \ \ 
(1 <\nu \le 2,\ \ \ 2 \le p < r\nu)\ .
\label{p.6}\end{equation}
The function $u$ has an angular derivative at infinity equal to
zero (i.e. $u(z)/z$ tends to 0 as $z \rightarrow \infty$ in non real
directions) because $u(z) = U(z)^{1/r}$, $U$ is anti-Herglotzian,
and $r>1$. Similarly $v$ and $\vhi$ 
have zero angular derivative at infinity.

\subsection{Analyticity}
\label{Ana}

The function $\vhi$ is holomorphic in $\Omega(-1/\lambda,\ \xi_{\max})$,
where $\xi_{\max} = \lambda^{-2}$ if $\chu(1/\lambda) \le 1/\lambda$ 
(as is the case for $\nu \le 1$, 
since $\lambda\chu(1/\lambda) \le z_1\lambda^{2-2\nu} <1$). 
In this case,
\begin{equation}
\vhi (\xi_{\max})\ =\ \vhi (\lambda^{-2})\ =\ 
z_1 v(\check u^{p-1}(\lambda^{-1}))
\ <\ {z_1 \over \lambda^\nu}\ < \lambda^{-2}\ .
\label{a.1}\end{equation}
If $\chu(1/\lambda) > 1/\lambda$, we denote $\xi_p = 1/\lambda^2$ and
$\lambda\xi_{p-1} = \chu^{-1}(1/\lambda)$. We construct by a descending
induction the strictly increasing sequence $\xi_1, \dots,\ \xi_p$ satisfying
$\chu^j(\lambda \xi_{p-j}) = \lambda \xi_p = 1/\lambda$. Supposing
$\xi_{p-j} < \dots \xi_p$ already constructed for a certain $j < p-1$,
we have $\chu^{j+1}(\lambda\xi_{p-j}) = \chu(1/\lambda) > 1/\lambda$,
while $\chu^{j+1}(\lambda) = \zeta_{j+1} < 1/\lambda$. Hence
$\lambda\xi_{p-j-1} = \chu^{-(j+1)}(1/\lambda)$ exists in $(1,\ \xi_{p-j})$.
We set $\xi_{\max} = \xi_{1}$ so that 
$\chu^{p-1}(\lambda\xi_{\max}) = 1/\lambda$.
Recalling that
$\chu^{p-1}(\zeta_1) = \lambda^{1-\nu}$, we find:
\begin{equation}
{\zeta_1 \over \lambda} \le 
\xi_{\max} = \xi_1 < \xi_2 < \dots < \xi_p = \lambda^{-2}\ .
\label{a.3}\end{equation}
The first inequality here is replaced by the equality
$\xi_{\max} = \zeta_1/\lambda$ when $\nu = 2$ (and, of course, $p >1$).
More generally $\chu^{p-j}(\zeta_j) = \lambda^{1-\nu}$ implies
$\zeta_j \le \lambda\xi_j$ for all $j \in [1,\ p-1]$, equality
holding when $\nu=2$.
Note (see (\ref{6.11}) and (\ref{6.13})) that 
$z_1/\lambda^{\nu-1} < \zeta_1 < 1/\lambda^{\nu-1}$, and
\begin{equation}
\vhi (\xi_{\max})\ =\ z_1 v(\lambda^{-1})
\ \le\ {z_1 \over \lambda^\nu}\ 
<\  \zeta_1/\lambda\ .
\label{a.4}\end{equation}

In both cases
the whole domain $\Omega(-1/\lambda,\ \xi_{\max})$ is a basin
of attraction of the fixed point 1 of $\vhi$, hence the domain of
$\psi$ is also $\Omega(-1/\lambda,\ \xi_{\max})$, and
\begin{eqnarray}
\psi(z) &=& {1 \over \lambda^{\nu r}} \psi(\vhi(z))\ ,\nonumber \\
\vhi(z) &=& \lambda^{\nu-1} \check u^p(\lambda z)\ ,
\label{a.5}\end{eqnarray}
hold for all $z \in \Omega(-1/\lambda,\ \xi_{\max})$. Also
\begin{equation}
u(z) = {1\over \lambda^{\nu}} u(\lambda^{\nu-1} \check u^p(\lambda z)),
\ \ \ \ z \in \Omega(-1/\lambda,\ 1)\ .
\label{a.6}\end{equation}

\subsection{Univalence for $p \ge 2$}

\label{univ}
We prove in this subsection that $\psi$ and $\vhi$ are univalent in 
$\Omega(-1/\lambda,\ \xi_{\max})$. We temporarily denote 
\begin{eqnarray}
\hphi_j(z) &=& -\check u^j(\lambda z), \ \ \ {\rm for}\ 0 \le j \le p-1\ ,
\nonumber \\
\hphi_p(z) &=& \vhi(z)\ .
\label{u.1}\end{eqnarray}
Let $c$ be fixed with $1<c < \min\{1/\lambda,\ \zeta_1/\lambda\}$. 
We first verify that each $\hphi_j$, $0\le j \le p$, 
maps the interval 
$(-1/\lambda,\ c)$ into an open interval $X_j$ with 
closure contained in $(-1/\lambda,\ c)$. 
This is clear in the case $j=p$, since $\hphi_p = \vhi$. 
For $j=0$, $\hphi_0(-1/\lambda) =1$, and 
$\hphi_0(c) = -\lambda c > -1$. 
If $1 \le j \le p-1$, $\hphi_j$ is decreasing,
$\hphi_j(c) < \hphi_j(-1/\lambda) \le 0$ and 
$\hphi_j(c) >  -\check u^j(\zeta_1) = -\zeta_{j+1} \ge -1/\lambda$. 
Let $X$ be the convex hull of $X_0 \cup\ \dots\ \cup X_p$. 
This is an open interval with closure contained in 
$(-1/\lambda,\ c)$, such that, for all $j = 0,\ \dots,\ p$, 
$\hphi_j((-1/\lambda,\ c)) \subset X$.

Suppose that $w'$ and $w''$ are distinct points in 
$\Omega(-1/\lambda,\ \xi_{\max})$ such that 
$\psi(w') = \psi(w'')$. This implies that $w'$ and $w''$ are 
not real, and have imaginary parts of the same sign. We inductively 
construct a sequence of triples 
$\{ w'_n,\ w''_n,\ j_n\}_{0\le n < \infty}$, where 
$w'_0 = w'$, $w''_0 = w''$, and, for all $n \ge 0$, 
$w'_n \ne w''_n$ are non-real, $\psi(w'_n) = \psi(w''_n)$,
and $0 \le j_n \le p$ is such that 
$w'_{n+1} = \hphi_{j_n}(w'_n)$, $w''_{n+1} = \hphi_{j_n}(w''_n)$.
Assuming that $w'_n$ and $w''_n$ have 
already been constructed, it follows from (\ref{a.5}) and the definition 
(\ref{u.1}) of the functions $\hphi_j$ that there is a unique $j_n$
in $[0,\ p]$ such that 
$\hphi_{j_n}(w'_n) \ne \hphi_{j_n}(w''_n)$ and 
either $j_n=p$ or 
$\hphi_{j_n+1}(w'_n) = \hphi_{j_n+1}(w''_n)$. This 
implies that 
$\psi(\hphi_{j_n}(w'_n)) = \psi(\hphi_{j_n}(w''_n))$, and
we take $w'_{n+1} = \hphi_{j_n}(w'_n)$, $w''_{n+1} = \hphi_{j_n}(w''_n)$.
It is easy to see (as e.g. in \cite{E2}) that, as $n$ tends to infinity,
the Poincar\'e distances, relative to $\bC_+ \cup \bC_-\cup X$,
of $w'_n$ and $w''_n$ to the segment $X$ tend to 0.
Therefore as $n$ becomes sufficiently 
large, the points $w'_n$ and $w''_n$ enter a complex neighborhood 
of the real segment $X$ so thin that $\psi$ is injective there, producing 
a contradiction. Thus $\psi$, and therefore also $u$ and $\vhi$ are 
univalent in their respective domains.

\subsection{Boundary values of $u$}
\label{bv}

We show in this subsection that the restriction of $u$ to the upper
half-plane $\bC_+$ extends to a continuous bounded injective function on the
closed upper half-plane $\ovl{\bC_+}$. The same, of course, holds in
the lower half-plane, since $u(z)=u^*(z^*)$.

We rewrite (\ref{a.6}) as
\begin{equation}
u(z) = F(u(-\lambda z)),
\label{bv.1}\end{equation}
\begin{equation}
F(z) = {1\over \lambda^{\nu}} u(\lambda^{\nu-1} \check u^{p-1}(z))\ .
\label{bv.2}\end{equation}
The function $F$ is anti-Herglotzian, holomorphic and univalent in 
$\Omega(-1,\ \zeta_1)$ and vanishes at $\zeta_1 = u(-\lambda)$.
It has a fixed point at $u(0) = z_1\lambda^{1-\nu}$ with 
$F'(z_1\lambda^{1-\nu}) = -1/\lambda$, and (since it is strictly decreasing)
no other fixed point in the real interval $[-1,\ \zeta_1]$.
The equation (\ref{bv.1}) can be rewritten as
$F = u\circ(-\lambda^{-1})\circ u^{-1}$ on the intersection of
the domain of $F$ with the range of $u$.
This range includes the real segment $(0,\ u(-\lambda^{-1}))$
and hence $(0,\ \zeta_1]$, since $u(-\lambda^{-1}) > u(-\lambda)= \zeta_1$.
Any periodic orbit of $F$ in $[-1,\ \zeta_1]$ 
must be contained in $(0,\ \zeta_1]$,
so that $\{u(0)\}$ is the only such orbit.
Therefore the Herglotz function $F^2$ also has $u(0)$ as its unique
real fixed point, with ${F^2}'(u(0)) = \lambda^{-2}$. Both $F$ and
$F^2$ have zero angular derivative at $\infty$ since 
$z\mapsto F(z)^r$ is anti-Herglotzian and $z\mapsto F^2(z)^r$ is
Herglotzian. 
As in [EL, E2], the theory of iterations
of maps of $\bC_+$ into itself 
(Wolff-Denjoy-Valiron Theorem, see [V, Mi]) shows that, 
uniformly on every compact
subset of $\bC_+$, $F^{2n}$ tends to a finite constant $c$ as 
$n \rightarrow \infty$. In particular for all $z$ in a fixed
compact subset of $\bC_-$, $u(\lambda^{-2n}z) = F^{2n}(u(z))$
tends uniformly to $c$, i.e. $c = u(-i\infty)$.

For $n = 0, 1, 2, ...$, we denote $I_n$ the closed real interval 
\begin{equation} 
I_n = (- \lambda )^{-n} [1,\  \lambda ^{-2}]. 
\label{bv.3}\end{equation}

We first consider the case when $\xi_{\max} = \lambda^{-2}$  
(in particular the case $0 < \nu \le 1)$, for which the argument of 
the case $p=1$ \cite{EL,E2} can be repeated almost verbatim. 
Let $z$ follow $I_0 -i0$. Then $u(z)$ follows the segment 
\begin{equation}
\tau_0 = e^{i\pi /r} [0,\ |U(\lambda ^{-2})|^{1/r}].
\label{bv.4}\end{equation}
If $z$ crosses the interior of $I_0$ into ${\bf C} _+$, $u$ gets continued 
by $v_0 \equiv e^{2\pi i/r} u$. 
The image of $\bCp$ given by $v_0$ is contained in 
$\{z\ :\ \pi/r < {\rm arg}\,z < 2\pi/r\}$. It is contained in $\bCp$
if and only if $r \ge 2$ (in particular if $r$ is an integer).
Let $\VV_0$ denote the open set
$\{z\in \bCp\ :\ v_0(z) \in \bCp\}$, and, for $n \ge 1$,
$\VV_n = -\lambda^{-1}\VV_{n-1}^*$ (so that $\VV_n = \bCp$ when $r \ge 2$).
If $z$ follows $I_1-i0$, then $-\lambda z$ follows 
$I_0+i0$ and, by (\ref{bv.1}), $u(z)$ follows the analytic arc 
$\tau_1 = F(\tau_0 ^*)$ which 
lies entirely in ${\bf C}_+$ except for its starting point,  
$u(-\lambda ^{-1}) = F(0)$. 
An easy induction shows that when $z$ follows $I_n -i0$, ($n \ge 1$),
$u(z)$ follows an analytic arc $\tau_n$ lying entirely in
${\bf C}_+$ for $n>1$, and $u$ can be continued across the interior of
$I_n$ into $\bC_+$ by a function $v_n$ holomorphic in $\VV_n$, with
$v_n(\VV_n) \subset \bCp$ and
\begin{eqnarray}
\tau_n &=& F(\tau_{n-1}^*)\ ,\nonumber \\
v_n(z) &=& F(v_{n-1}(-\lambda z^*)^*)\ .
\label{bv.5}\end{eqnarray}
The starting point of $\tau_{n+2}$ is the end of $\tau_n$, and
$\tau_{n+2} = F^2(\tau_n)$. Hence the arcs $\tau_n$ tend 
to the point $c$. Thus $u|\bC_+$ extends to a continuous bounded
function on $\ovl{\bC_+}$. This function is
injective. Indeed at each step of its inductive construction,
a new extension is obtained by composing copies of the previously
constructed extension and scalars. 

We now consider the case when $\xi_{\max} < \lambda^{-2}$ which occurs
if $\check u(\lambda^{-1}) > \lambda^{-1}$. (In particular 
for $\nu=2$, $\zeta_p = \lambda^{-1} < \zeta_{p+1} = \chu(\lambda^{-1})$.) 
Recall that we denote $\xi_j$,
for $1 \le j \le p$, the unique number in 
$[\zeta_1/\lambda,\ \lambda^{-2}]$ such that
\begin{equation}
\check u^{p-j}\,(\lambda\,\xi_j) = \lambda^{-1}\ ,
\label{bv.6}\end{equation}
and that
\begin{displaymath}
\xi_{\max} = \xi_1 < \cdots <  \xi_p = \lambda^{-2}\ ,
\end{displaymath}
\begin{equation}
\zeta_{j}\ \le\ \lambda\xi_j\ \ \ \forall j \in [1,\ p]\ .
\label{bv.7}\end{equation}
Suppose $z$ follows $[1,\ \xi_{\max}] \mp i0$. Then 
$u(z) = e^{\pm i\pi/r}|U(z)|^{1/r}$ follows the segment 
$e^{\pm i\pi/r}[0,\ |U(\xi_{\max})|^{1/r}]$. Hence if $z$ follows
$(-1/\lambda)[1,\ \xi_{\max}]-i0$, $u(z)$ is given by (\ref{bv.1}),
and follows an analytic arc entirely contained in $\bC_+$ except
for its starting point $\chu(1/\lambda)$. Thus $z\mapsto u(z-i0)$
now has a continuous, non-real extension to 
$[-\xi_{\max}/\lambda,\ -1/\lambda)$.
The extension thus obtained of
$u|\bC_-$ to $\bC_-\cup [-\xi_{1}/\lambda,\ \xi_{1}]$ is also
obviously injective, as well as the conjugate extension of
$u|\bC_+$. 
Recall also that 
\begin{equation}
1/\lambda < \chu(1/\lambda) \le z_1\lambda^{1-2\nu}
< \zeta_1/\lambda^\nu \le \xi_{\max}/\lambda\ .
\label{bv.7b}\end{equation}
Hence
\begin{equation}
\chu(1/\lambda) \in (1/\lambda,\ \xi_{\max}/\lambda)\ ,
\label{bv.7a}\end{equation}
and
\begin{equation}
1\le \lambda^{\nu-2} < \lambda^{\nu-1}\chu(1/\lambda) < \zeta_1/\lambda\ .
\label{bv.7c}\end{equation}
This shows that $\lambda^{\nu-1}\chu(1/\lambda)$ is in the domain
of analyticity of $U$ and $U$ is negative there.

We assume inductively that there exists, for a certain $j \in [1,\ p-1]$,
a continuous injective 
extension, temporarily denoted $u_{(j)}$,
of $u|\bC_-$ to $\bC_-\cup [-\xi_j/\lambda,\ \xi_j]$,
such that $u_{(j)}(z-i0) \in \bC_+$ if $z \in [-\xi_j/\lambda,\ -1/\lambda)$
or if $z \in (1,\ \xi_j]$. This implies of course a symmetrical situation
for $u|\bC_+$. By abuse of notation we also denote 
$u_{(j)}(z+i0) = u_{(j)}(z^*-i0)^*$.
We also assume that (\ref{a.6}) holds with $u$ replaced by $u_{(j)}$
in the domain of the latter.
In order to prove the same for $j+1$, we denote
\begin{equation}
u_{(j+1/2)}(z\mp i0) = 
\lambda^{-\nu} u_{(j)}(\lambda^{\nu-1}\check u_{(j)}^{j}
(\check u_{(j)}^{p-j}(\lambda z\mp i0)))\ .
\label{bv.7d}\end{equation}
Note that the rhs of the above equation is equal to $u_{(j)}(z\mp i0)$ in the
domain of $u_{(j)}$ by the induction hypothesis.
Thus $u_{(j+1/2)}$ is a new extension of $u|\bC_-$, 
which is injective wherever it is defined. 
If $z$ increases along $(\xi_j,\ \xi_{j+1}]$, 
$\chu_{(j)}^{p-j}(\lambda z-i0)$ moves along 
$(1/\lambda,\ \chu(1/\lambda)]-i0$.
This, by the induction hypothesis and (\ref{bv.7a}), is within the
domain of the already constructed $\chu_{(j)}$, and
$\chu_{(j)}^{j}(\chu_{(j)}^{p-j}(\lambda z-i0))$ moves along an arc entirely
contained in $\bC_-$, 
and $u_{(j+1/2)}(z-i0)$ moves along an arc entirely contained 
in $\bC_+$. A little more detail is needed, since
$\chu(1/\lambda)$ is real, when $z$ moves in a small
real interval containing $\xi_j$ so that $\chu^{p-j}(\lambda z-i0)$ 
moves along a small real interval containing $1/\lambda$.
If $j=1$, then $u_{(1)}$ is continuous and non-real at 
$\lambda^{\nu-1}\chu(1/\lambda) \pm i0$ since, as noted above,
this is a point of analyticity of $U$.
If $j >1$, $\chu_{(j)}^2(1/\lambda \pm i0)$ is defined and
non-real by (\ref{bv.7a}). Denote now 
$u_{(j+1)}(z - i0) = 
\lambda^{-\nu} u_{(j+1/2)}(\lambda^{\nu-1}\chu_{(j+1/2)}^p(\lambda z-i0))$.
This is a continuous injective extension of $u|\bCm$ to
$\bCm \cup [-\xi_{j+1}/\lambda,\ \xi_{j+1}]$ since the arc
$u_{(j+1/2)}([\xi_j,\ \xi_{j+1}]\mp i0)$ is entirely contained in $\bCpm$.
The construction makes it obvious that (\ref{a.6}) holds with
$u$ replaced by $u_{(j+1)}$ in the domain of the latter.

We conclude that $u|\bC_-$ has a continuous injective
extension to $\bC_- \cup [-\lambda^{-3},\ \lambda^{-2}]$ which takes real
values only on $[-1/\lambda,\ 1]$. It maps $I_0 -i0$ onto a union
$\tau_0$ of $p$ consecutive arcs contained in $\bC_+$ except 
for the point $u(1)=0$ :
$\tau_0 = \tau_{00}\cup\dots\tau_{0(p-1)}$, with 
$\tau_{00} = u([1,\ \xi_1]-i0)$ and
$\tau_{0j} = u([\xi_{j},\ \xi_{j+1}]-i0)$ for $1\le j < p$.
It maps $I_1 - i0$ onto another finite union 
$\tau_1 = \tau_{10}\cup\dots\tau_{1(p-1)}$, with 
$\tau_{1j} = F(\tau_{0j}^*)$,
contained in $\bC_+$ except for the point $u(-1/\lambda)$.
As in the previous case, $u|\bC_-$ extends to a continuous injective
function on~$\ovl{\bC_-}$. The images $\tau_n = u(I_n-i0)$ all lie in $\bC_+$
for $n > 1$. The sequence of the $\tau_n = F(\tau_{n-1}^*) = F^2(\tau_{n-2})$
tends to the point $c$. 

Let $I_{00} = [1,\ \xi_1]$, $I_{0j}= [\xi_{j},\ \xi_{j+1}]$
for $1\le j < p$,
and $I_{nj} = (-\lambda)^{-n}I_{0j}$ for $n\in\bN$, $0 \le j <p$.
If $z$ crosses $(1,\ \xi_{\max})$ from 
$\bCm$ into $\bCp$, $u(z)$ gets
continued by $v_{00}(z) = e^{2i\pi/r}u(z)$. If $z$ crosses 
$(-\xi_{\max}/\lambda,\ -1/\lambda)$ from $\bCm$ into $\bCp$, $u(z)$ gets
continued by $v_{10}(z) = F(v_{00}(-\lambda z^*)^*)$, holomorphic
in $\VV_{10} = \VV_1$. If $z$ crosses $(\xi_j,\ \xi_{j+1})$ 
from $\bCm$ into $\bCp$ (with $1\le j <p$), $u(z)$ gets continued by
$v_{0j}(z) = \lambda^{-\nu}u(\lambda^{\nu-1}\chu^{j-1}
(v_{10}(-\chu^{p-j}(\lambda z^*))^*))$. 
If $z$ crosses $(-\xi_{j+1}/\lambda,\ -\xi_{j}/\lambda)$ 
from $\bCm$ into $\bCp$, $u(z)$ gets continued by
$v_{1j}(z) = F(v_{0j}(-\lambda z^*)^*)$. If $z$ crosses the interior
of $I_{nj}$ from $\bCm$ into $\bCp$, $u(z)$ gets continued by
$v_{nj}(z) = F(v_{(n-1)j}(-\lambda z^*)^*)$. If $r \ge 2$, all the
functions $v_{nj}$ are holomorphic in $\bCp$ and map it into itself.

Note that, in all cases, the extension of 
$u$ to $\bbCm$ (resp. $\bbCp$) takes real values only on $[-\lambda^{-1},\ 1]$.
The function $F|\bC_+$ (resp. $F|\bC_-$) also has a continuous injective
extension to $\bbCp$ (resp. $\bbCm$) which takes real values only on
the real segment $[-1,\ \zeta_1]$. The point $c$ cannot be real. Indeed
if we suppose it is and let $w_0 \in \bC_+$, $w_n = F^{2n}(w_0)$ for
$n \in \bN$, the sequences $\{w_n\}$ and $\{F^2(w_n)\}$ both tend to $c$,
so that, by the continuity of the extensions of $F$ to $\bbCpm$,
$F^2(c+i0) = c$. Since this is real, $F(c+i0)$, hence also $c$, must belong 
to $[-1,\ \zeta_1]$, and $c$ is a fixed
point of $F^2$, i.e. $c$ coincides with $z_1\lambda^{1-\nu}$. But the
latter is repulsive, contradicting the attractive property of $c$.
Hence $c$ is in $\bC_+$ and is a fixed point of $F^2$. It is attractive
and unique by Schwarz's lemma applied to $F^2|\bC_+$. Therefore the
compact sets $\tau_n$ converge geometrically to $c$. It follows that the
functions $u$, $\vhi$, $\psi$ are all bounded.

\subsection{Commutativity for $\nu = 2$}

\label{commut}
The following is a transcription into the notations of this paper of
(a special case of) a result due to
O.~Lanford. This will prove that the properties of $u$ and $\psi$  
recalled at the beginning of this section suffice to imply, in the case 
$\nu =2$, a form of commutativity for the functions $\xi$ and $\eta$ 
given by
\begin{equation}
\xi = (-u)^{-1},\ \ \ \eta = -\lambda \xi \circ (-1/\lambda),\ \ 
\eta^{-1} = \lambda \check u \circ (1/\lambda)\ .
\label{C.1}\end{equation}
Recall that the functional equations (\ref{a.5}), (\ref{a.6}),
and $\nu =2$, imply that
$\xi$ and $\eta$ satisfy the system (\ref{1.5}).
With the notations of the beginning of this section, we have

\begin{lemma}
\label{com} 
{\rm (Lanford)} For every solution with $\nu=2$, 
\begin{equation}
\psi(\lambda u(-z/\lambda)) =
-\lambda^r \psi(u(z)/\lambda)\ \ \ 
{\rm for\ all\ } z \in \Omega(-\lambda,\ 1)\ .
\label{C.6}\end{equation}
\end{lemma}

\PROOF
The domains of the two anti-Herglotzian functions 
\begin{eqnarray}
F_1 &=& \psi \circ \lambda \circ u \circ (-1/\lambda)\ ,\nonumber \\
F_2 &=& -\lambda^r \psi \circ (1/\lambda)\circ u\ ,
\label{C.2}\end{eqnarray}
are equal to 
$\Omega(-\lambda,\ 1)$. Indeed the function 
$z \mapsto \lambda u(-z/\lambda)$ has this domain and maps $-\lambda$ to 0, 
and 1 to $z_1 v(1/\lambda) \le z_1/\lambda^2 < \zeta_1/\lambda$, hence
it maps $\Omega(-\lambda,\ 1)$ into the domain of $\psi$.
The function $(1/\lambda) u$ is holomorphic in 
$\Omega(-1/\lambda,\ 1)$. It maps 1 to 0 and $-\lambda$ to 
$\zeta_1/\lambda$, hence it also maps $\Omega(-\lambda,\ 1)$ into the domain 
of $\psi$ (and $F_2$ has a branch point at $-\lambda$ since $\psi$ has one at
$\zeta_1/\lambda$). 
We now substitute for $u$, in the equation for $F_1$, the r.h.s. of
(\ref{a.6}), and substitute for $\psi$, in the equation for $F_2$, 
the r.h.s. of the first equation in (\ref{a.5}). This gives 
\begin{equation}
F_1 = -{1\over \lambda^r} F_2 \circ G_0\ ,\ \ \ \ 
F_2 = -{1\over \lambda^r} F_1 \circ G_0\ ,
\label{C.3}\end{equation}
\begin{equation}
G_0(z) = \lambda \check u^p(-z) = \vhi(-z/\lambda)\ .
\label{C.3.1}\end{equation}
Since the functional equations (\ref{a.6}) and (\ref{a.5}) hold with 
domains, so does the system (\ref{C.3}). In fact 
the anti-Herglotzian function $G_0$ maps the domain
$\Omega(-\lambda,\ 1)$ into itself, since $G_0$ is holomorphic in 
$\Omega(-\zeta_1,\ 1)$ and satisfies: 
\begin{equation}
G_0(1) = \vhi(-1/\lambda) \ge 0\ ,\ \ \ G_0(0) = \vhi(0) < 1,\ \ \ 
G_0(-\lambda) = \vhi(1) = 1\ .
\label{C.4}\end{equation}
Since $G_0$ is strictly decreasing on $[-\lambda,\ 1]$, it has there a unique 
fixed point $\bar x \in (0,\ 1)$ which, by Schwarz's lemma, is attractive 
and has $\Omega(-\lambda,\ 1)$ as a basin of attraction. 
Let $\kappa = -G_0'(\bar x) \in (0,\ 1)$, and let $h$ be the linearizer of 
$G_0$ at $\bar x$, normalized by $h'(\bar x) =1$. This is a function 
holomorphic in $\Omega(-\lambda,\ 1)$, and satisfying, in this domain,
$h= (-1/\kappa)h\circ G_0$ (in particular $h(\bar x)=0$). 
The point $\bar x$ is also the unique fixed 
point of the function $G_0^2$ in $\Omega(-\lambda,\ 1)$ and its 
normalized linearizer is also $h$ . On the other 
hand, because $G_0$ maps $\Omega(-\lambda,\ 1)$ into itself, the equation 
obtained by substituting the second equation in (\ref{C.3}) into the first, 
\begin{equation}
F_1 = {1\over\lambda^{2r}} F_1 \circ G_0^2
\label{C.5}\end{equation}
holds in $\Omega(-\lambda,\ 1)$. Therefore 
$F_1 = c_1 h$ with $c_1 = F'_1(\bar x)$, and $\kappa = \lambda^r$. 
The second equation in (\ref{C.3}) now reads 
$F_2 = (-c_1/\kappa)h\circ G_0 = c_1 h$. Therefore $F_1$ and $F_2$ coincide,
which is the assertion of the lemma.

In particular, for $z= -\lambda$, this gives 
$1 = \psi(0) = -\lambda^r \psi(\zeta_1/\lambda)$, i.e.
$\psi(\zeta_1/\lambda) = -1/\lambda^r$. 
Both sides of (\ref{C.6}) must vanish at $\bar x$, hence 
$\check u(\bar x/\lambda) = 1/\lambda$, $u(\bar x) = \lambda$ so that
$\bar x = \lambda \zeta_{p-1} = -\zeta_{-1}$. 
Since $F_2(1) = -\lambda^r$, the 
common range of $F_1(z)$ and $F_2(z)$ as $z$ varies in 
$[-\lambda,\ 1]$ is $[-\lambda^r,\ 1]$.
The identity (\ref{C.6}) continues to 
hold if $\psi$ is replaced with $U = (z_1/\lambda)^r \psi$ on both sides. 
In order to translate this identity in terms of $\xi$ and $\eta$, 
we denote
\begin{equation}
q(z) = |z|^r {\rm sign}(z) \ \ \forall z \in \bR,\ \ \
\wh u(z) = q^{-1}\circ U(z)\ \ \forall z \in
[-1/\lambda,\ \zeta_1/\lambda].
\label{C.7}\end{equation}
The function $\wh u$ is strictly decreasing, with range containing 
$[-z_1/\lambda^2,\ 1/\lambda]$, coincides with $u$ on
$[-1/\lambda,\ 1]$, and satisfies
\begin{equation}
\matrix{
&\wh u = (1/\lambda^2)\, \wh u\circ \lambda \check u^p 
\circ \lambda \ \ \  &{\rm on\ } [-1/\lambda,\ \zeta_1/\lambda]\ ,
\hfill \cr
&\wh u \circ \lambda u \circ (-1/\lambda) \ =\ 
-\lambda \wh u \circ (1/\lambda)\, u \ \ \ &{\rm on\ } 
[-\lambda,\ 1]\ .\hfill \cr}
\label{C.8}\end{equation}
Let 
\begin{equation}
\wh \xi = (-\wh u)^{-1},\ \ \ \ 
\wh \eta = -\lambda \wh \xi \circ (-1/\lambda)\ .
\label{C.9}\end{equation}
Then $\wh \xi$ is an extension of $\xi$ to an interval containing 
$(-1/\lambda,\ z_1/\lambda^2)$, $\wh \eta$ is an extension of $\eta$,  
and 
\begin{equation}
\matrix{
&\wh \xi = (1/\lambda^2)\, \eta^p \circ \wh \xi \circ \lambda^2\ ,
\ \ \ \wh \xi = (-1/\lambda)\, \wh \eta \circ (-\lambda)\ ,
&{\rm on\ } (-1/\lambda,\ z_1/\lambda^2)\ ,\hfill\cr
&\eta \circ \wh \xi = \xi \circ \wh \eta 
&{\rm on\ } (-z_1/\lambda,\ z_1)\ .\hfill\cr}
\label{C.10}\end{equation}

\section{Behavior of fixed points as $r \rightarrow \infty$, $0<\nu\le 1$}

\label{larger}
In this section we consider only the cases when $0<\nu\le 1$
and $p \ge 2$ (and, of course, $r\nu - 1-(p-1)(1-\nu) >0$).
In the case $p = 1$, the behavior of solutions as $r \rightarrow \infty$
was first elucidated by Eckmann and Wittwer in \cite{EW}, and also
studied in \cite{E1} (for $\nu = 1$) and \cite{EE} 
(for $1 \le\nu \le 2$),
and the method of \cite{E1,EE} extends trivially to $0 < \nu \le 1$.
The case $p \ge 2$ requires some additional work.

\subsection{The functions $V$ and $W$}

\label{VW}
The functional equation implies
\begin{equation}
\psi(z) = V(\psi(-\lambda z)) = W(\psi(\lambda^2 z)),\ \ \ \ \ 
\forall z \in \bC_+\cup\bC_-\cup (-1/\lambda,\ 1/\lambda^2)\ ,
\label{v1}\end{equation}
where
\begin{eqnarray}
V(\zeta) &=& f(\zeta^{1/r}),\nonumber \\
f(z) &=& {1 \over \lambda^{r\nu}}
\psi(\lambda^{\nu -1}\, \check u^{p-1}(z_1\lambda^{1-\nu}\,z))\ ,
\nonumber \\
W &=& V \circ V\ .
\label{v2}\end{eqnarray}
Recall that
\begin{equation}
\check u(z) = z_1\lambda^{1-\nu}\,\psi(-z)^{1/r}\ .
\label{v3}\end{equation}
The function $f$ is anti-Herglotzian and holomorphic in
$\Omega(-1/z_1\lambda^{1-\nu},\ 1/z_1\lambda^{2-\nu})$.
We denote 
$\zeta_{\rm max} = (1/z_1\lambda^{2-\nu})^r$.
These functions satisfy
\begin{equation}
V(1) = f(1) =1,\ \ \ \ 
V'(1) = -{1 \over\lambda},\ \ \ \ f'(1) = -{r \over\lambda}\ .
\label{v4}\end{equation}
Since
$\psi(1) = 0$, $V$ vanishes at $\alpha = \psi(-\lambda)$, and 
$f$ vanishes at 
$v(\lambda) = (z_1\lambda^{1-\nu})^{-1}\zeta_1$.
We also define 
\begin{equation}
\wh V(\zeta) = 1-V(1-\zeta),\ \ \ \ 
\wh W = \wh V\circ \wh V\ .
\label{v5}\end{equation}
Since the functional equations (\ref{v1})
hold for all $z$ in the domain of $\psi$, 
the real ranges of $V$ and $W$ contain that of $\psi$.
The following estimates follow \cite{EE} and \cite{E1}.
In the domain of $V$,
\begin{equation}
-{V''(\zeta) \over V'(\zeta)} = 
{1 \over r\zeta}\left (
r - 1 - {z\,f''(z) \over f'(z)} \right ),\ \ \ \ 
z = \zeta^{1/r}\ .
\label{v6}\end{equation}
For real $\zeta \in (0,\ \zeta_{\rm max})$,
\begin{eqnarray}
-{V''(\zeta) \over V'(\zeta)} &\ge&
{1 \over r\zeta}\left (
r - 1 - {2z \over 1/\lambda^{2-\nu}z_1-z} \right )\nonumber \\
&=& {1 \over r\zeta}\left (
r - {1 + \lambda^{2-\nu}z_1\,z \over 1 - \lambda^{2-\nu}z_1\,z}
\right )\ .
\label{v6.1}\end{eqnarray}
Recalling the bound $r\nu \ge (1+\lambda)/(1-\lambda)$, we find that
\begin{equation}
-{V''(\zeta) \over V'(\zeta)} \ge {1-\nu \over \zeta}\ \ \ \ {\rm for\ }
0 < \zeta \le (z_1\lambda^{1-\nu})^{-r}
\label{v6.2}\end{equation}
This is in particular satisfied if 
$\zeta = \alpha = ((z_1\lambda^{1-\nu})^{-1}\zeta_1)^r$, since
$\zeta_1 \le \lambda^{1-\nu} \le 1$.
Integrating the inequality (\ref{v6.2}) from 1 to $\zeta >1$, using 
$V'(1) = -1/\lambda$ and $V(1) = 1$, gives
\begin{equation}
V(\zeta) > 1 - {1\over \lambda\nu}(\zeta^\nu -1)\ \ \ \ 
\Rightarrow \alpha > (1+\lambda\nu)^{1/\nu} \ge (1+\lambda).
\label{v6.3}\end{equation}
It follows similarly from (\ref{v6}) that
\begin{equation}
-{V''(\zeta) \over V'(\zeta)} \le
{1 \over r\zeta}\left (
r - 1 + {2z \over z + \lambda^{\nu-1}/z_1} \right )\ \ \ 
= {1 \over \zeta}\left (
1 - {1- z_1\lambda^{1-\nu}\,z \over r(1+ z_1\lambda^{1-\nu}\,z)}\right )\ ,
\label{v7}\end{equation}
so that
\begin{equation}
-{V''(\zeta) \over V'(\zeta)} \le {1 \over \zeta}
\ \ \ \ \forall \zeta \in (0,\ \alpha)\ .
\label{v8}\end{equation}
Integrating this from 1 to $\zeta > 1$ gives
\begin{equation}
-V'(\zeta) \ge {1 \over \lambda\zeta}
\ \ \ \ \forall \zeta \in (1,\ \alpha)\ ,
\label{v8.1}\end{equation}
\begin{equation}
V(\zeta) \le 1 - {1 \over \lambda}\log \zeta
\ \ \ \ \forall \zeta \in (1,\ \alpha)\ \ \ \
\Rightarrow \alpha \le e^\lambda\ .
\label{v8.2}\end{equation}
Since $V = f \circ q^{-1}$ where $q^{-1}(\zeta) = \zeta^{1/r}$,
the Schwarzian derivative $SV$ of $V$ satisfies, for real
$\zeta$ in the domain of $V$,
\begin{equation}
SV(\zeta) \ge Sq^{-1}(\zeta) = 
{1 - r^{-2}\over 2\zeta^2}\ .
\label{v9}\end{equation}
The function $W$ is Herglotzian and holomorphic in
$\Omega(0,\ \alpha)$, where $\alpha = \psi(-\lambda) = V^{-1}(0)$ 
(since $V(0) \le \lambda^{-r\nu} < \zeta_{\max}$).
It has a repelling fixed point at 1 with
multiplier $\lambda^{-2}$. $\wh W$ is Herglotzian and holomorphic in
$\Omega(1-\alpha,\ 1)$ and has a fixed point at 0. By (\ref{v9}),
\begin{equation}
SW( \zeta )  \ge  {(1- r^{-2} ) \over 2}
\left (  {V '( \zeta )^{2} \over V( \zeta )^{2} } +
{1 \over \zeta^{2} }\right )\ .
\label{v9.1}\end{equation}

\subsection*{Lower bound for $\wh W$ in $[0,\ 1]$.}

For $ 0 < \zeta < 1 $, the convexity of $V$ 
implies: 
\begin{equation} 
- V '( \zeta )  \ge   {V( \zeta ) -1 \over 1 - \zeta } , 
\label{v9.1a}\end{equation} 
hence 
\begin{equation}
-{V( \zeta ) \over V ' ( \zeta )} \le 
1 - \zeta - {1 \over V ' ( \zeta )} \le 1 - \zeta + \lambda\ ,
\label{v9.1b}\end{equation}
It follows that 
\begin{equation} 
2SW( \zeta ) \ge   (1- r^{-2} ) \left [ {1 \over 
( 1 - \zeta + \lambda )^{2} } + {1 \over 
 \zeta^{2} }  \right ] , 
\label{v9.1c}\end{equation} 
and hence 
\begin{equation} 
2S\widehat W ( \zeta )  \ge   (1- r^{-2} ) \left [ {1 \over 
 ( \zeta + \lambda )^{2} } + {1 \over 
(1- \zeta )^{2} }  \right ] . 
\label{v9.2}\end{equation} 
In (0, 1), the r.h.s. has a minimum at $ \zeta = (1- \lambda )/2$, 
and, using the bound on $r \ge (1+\lambda)/(1-\lambda)$, we get 
\begin{equation}
{d \over d \zeta }\,  
{{\widehat W } ''( \zeta ) \over 
{\widehat W } '( \zeta ) } \ge  
S\widehat W ( \zeta )  \ge   s( \lambda )  \equiv   
{ 16 \lambda \over (1+ \lambda )^{4} } . 
\label{v9.3}\end{equation}
By (\ref{v8}) and
\begin{equation}
{\wh W''(\zeta) \over \wh W'(\zeta)} =
{\wh V''(\wh V(\zeta)) \over \wh V'(\wh V(\zeta))}\wh V'(\zeta) +
{\wh V''(\zeta) \over \wh V'(\zeta)}\ ,
\label{v10}\end{equation}
it follows that
\begin{equation}
{\wh W''(0) \over \wh W'(0)} =
-\left({1\over \lambda} -1\right) {\wh V''(0) \over \wh V'(0)} =
\left({1\over \lambda} -1\right) {V''(1) \over V'(1)} \ge
-\left({1\over \lambda} -1\right)\ .
\label{v11}\end{equation}
Hence, 
\begin{equation}
{\wh W''(\zeta) \over \wh W'(\zeta)} \ge
{\wh W''(0) \over \wh W'(0)} +s(\lambda) \zeta
\ge -\left({1\over \lambda} -1\right)
+ s(\lambda) \zeta\ ,
\label{v13}\end{equation}
\begin{eqnarray}
\log \wh W'(\zeta) &\ge& 2\log (1/\lambda) 
-\left({1\over \lambda} -1\right)\zeta
+s(\lambda)\zeta^2/2  \nonumber \\
&\ge& \left [ 2\log (1/\lambda) 
-\left({1\over \lambda} -1\right) \right]
+ s(\lambda)\zeta^2/2 \ .
\label{v14}\end{eqnarray}
As a function of $\lambda$ in $(0,\ 1)$, the first bracket in the last
expression has a unique maximum at $1/2$ and vanishes at 1. Since it is
positive at $e^{-1}$, it is non negative in $[e^{-1},\ 1]$.
Hence, for $\lambda \ge e^{-1}$ and $0\le \zeta < 1$,
\begin{equation}
\wh W'(\zeta) \ge 1+s(\lambda)\zeta^2/2 \ ,
\label{v15}\end{equation}
\begin{equation}
\wh W(\zeta) \ge \zeta \left(1 + {s(\lambda)\over 6}\zeta^2 \right)\ ,
\label{v16}\end{equation}
and we note that, for $\lambda \ge 1/4$, $s(\lambda) \ge 1$.

On the other hand $\wh W$ is Pick with 0 angular derivative
at infinity in $\bC_+\cup\bC_-\cup (1-\alpha,\ 1)$,
and vanishes at 0.
Hence there is a positive measure $\rho$ with support in
$\bR \setminus (1-\alpha,\ 1)$ such that
\begin{equation}
\wh W(\zeta) =
\int_{\bR \setminus (1-\alpha,\ 1)}
\left ( {1\over t - \zeta} - {1\over t} \right ) d\rho(t) \ ,\ \ \ \
\int_{\bR \setminus (1-\alpha,\ 1)}
{d\rho(t) \over t^2} 
= {1 \over \lambda^2}\ .
\label{v16.1}\end{equation}
Hence, for $0 \le \zeta <1$,
\begin{eqnarray}
\wh W(\zeta) \ge {\zeta \over \lambda^2}\ 
\inf_{t \notin (1-\alpha,\ 1)}\ 
{t \over t-\zeta}
&=& {\zeta (\alpha -1) \over \lambda^2(\alpha -1 + \zeta)}\nonumber \\
&\ge&
{\zeta \over \lambda(1 +\lambda)}\ .
\label{v16.2}\end{eqnarray}
Here we have used the lower bound (\ref{4.17}) for $\alpha$.
For $\lambda \le 1/2$, this implies
$\wh W(\zeta) \ge 4\zeta/3 \ge \zeta(1+\zeta^2/6)$, so that,
for all $\lambda$ and all $\zeta \in (0,\ 1)$,
\begin{equation}
\wh W(\zeta) \ge \zeta(1+ c'\zeta^2),\ \ \ \ \ c' = 1/6.
\label{v16.3}\end{equation}

\begin{remark}
\label{invineq}
Let $\zeta$, $y$, $a'$, and $m$ be strictly positive real numbers such that
\begin{equation}
0\ \ <\ \ \zeta(1+a'\zeta^2)\ \ \le\ \ y\ \ \le\ \ m\ .
\label{v16.4}\end{equation}
Then
\begin{equation} 
\zeta \le y (1- ay^2),\ \ \ \ \ \ \ a = {a' \over 1+ 3 a' m^2}\ .
\label{v16.5}\end{equation}
\end{remark}
Indeed, note first that $am^2 \le 1/3 < 1$. Moreover 
$\zeta \le z$ for any $z$ such that 
$a'z^3 + z -y \ge 0$, and inserting $z = y(1-ay^2)$ in this expression gives
\begin{displaymath}
y^3 [a'(1-ay^2)^3 -a] \ \ \ge\ \  y^3 [a'(1- 3 am^2) -a] = 0\ .
\end{displaymath}

\noindent
This remark (with $m=1$) and the lower bound (\ref{v16.3})
imply that $\wh W^{-1}$ is defined on 
$[0,\ 1)$, and that, for all 
$y \in [0,\ 1)$,
\begin{equation}
\wh W^{-1}(y) \le y(1 - cy^2),\ \ \ \ 
c = {c' \over 1 +3 c'} = 1/9\ .
\label{v17}\end{equation}

\subsection*{Lower bound for $W$ in $[1,\ \alpha]$.}

For $1 \le \zeta \le \alpha$, the inequalities (\ref{v9.1}) and (\ref{v8.1}),
together with $0 \le V(\zeta) \le 1$, give
\begin{equation}
SW(\zeta) \ge {1 \over 2\zeta^2}(1-r^{-2})(\lambda^{-2} +1)
\ge {1 \over \zeta^2}{2 \over (1+\lambda)^2}(\lambda^{-1} + \lambda)\ .
\label{v18}\end{equation}
The last inequality follows from the lower bound on $r$ already used above.
The last expression is decreasing in $\lambda$, so that, finally,
\begin{equation}
SW(\zeta) \ge {1 \over \zeta^2}
\ \ \ \ \forall \zeta \in (1,\ \alpha)\ .
\label{v19}\end{equation}
Since $W''(1)/W'(1) = -\wh W''(0) /\wh W'(0) \ge 0$ (see (\ref{v11})),
for $1 \le \zeta \le \alpha$,
\begin{equation}
{W''(\zeta) \over W'(\zeta)} \ge
\int_1^\zeta \, t^{-2}dt = (\zeta-1)/\zeta \ge (\zeta-1)/e\ ,
\label{v20}\end{equation}
by using the upper bound $\alpha \le e$, and hence
\begin{eqnarray}
W'(\zeta) &\ge& \lambda^{-2}(1+ (\zeta-1)^2/2e)\ ,\nonumber \\
W(\zeta) -1 &\ge& (\zeta-1)( 1+ k'(\zeta-1)^2),\ \ \ \ k'= 1/6e
\ \ \ \ \forall \zeta \in (1,\ \alpha)\ .
\label{v21}\end{eqnarray}
The function $\underline W(\zeta) = W(\zeta+1) -1$ is thus defined on
$[0, \alpha -1]$, where it satisfies
\begin{equation}
\underline W(\zeta) \ge \zeta (1+ k'\zeta^2)\ .
\label{v22}\end{equation}
We note that $W(\alpha) = W(\psi(-\lambda)) = \psi(-\lambda^{-1})$,
hence the range of $W|(1,\ \alpha)$ contains in particular 
$\psi(-1)$. We wish to apply 
Remark \ref{invineq} to the inverse
function $\unl W^{-1}$ restricted to $[0,\ \psi(-1)-1]$, and we first
obtain an upper bound for $\psi(-1)$. We use the representation (\ref{2.4}) :
\begin{equation}
\matrix{
\log \psi(-1) - \log \psi(-\lambda) &=&
\displaystyle \int_{\bR \setminus (-\lambda^{-1},\ 1)}
\sigma(t) \left ( 
{ \displaystyle 1 \over \displaystyle \strut t+\lambda} - 
{ \displaystyle 1 \over \displaystyle \strut t+1} \right )\,dt&\cr
&\le& \displaystyle \int_{\bR \setminus (-\lambda^{-1},\ 1)}
\left ( 
{ \displaystyle 1 \over \displaystyle \strut t+\lambda} - 
{ \displaystyle 1 \over \displaystyle \strut t+1} \right )\,dt&
=\  \log 2 \hfill\cr}
\label{v23}\end{equation}
which yields (using (\ref{v8.2}))
\begin{equation}
\psi(-1) \le 2 \psi(-\lambda) \le 2e^\lambda < 2e\ .
\label{v24}\end{equation}
Thus (\ref{v22}) and 
Remark \ref{invineq}, with $m= 2e-1$, show that
\begin{equation}
\unl W^{-1}(y) \le y(1- ky^2),\ \ \ \ \ \ 
k = 1/(6e + 3(2e -1)^2),\ \ \ \ \ \ 
\forall y \in [0,\ \psi(-1) -1]\ .
\label{v25}\end{equation}
Note that we have obtained the following bounds:
\begin{equation}
1+\lambda \le \psi(-\lambda) \le e^{\lambda},\ \ \ \ \ 
\psi(-1) \le 2 \psi(-\lambda) \le 2e^{\lambda}\ .
\label{v26}\end{equation}
This provides upper and lower bounds for $y_0 = z_1^r$. Indeed
from $\zeta_1 = z_1 \lambda^{1-\nu}\,v(\lambda) \le \lambda^{1-\nu} =
\zeta_p$, and 
$z_1 \lambda^{1-\nu}\,v(1) \ge z_1 \lambda^{1-\nu}\,v(\zeta_p) \ge
\zeta_p$, it follows $z_1 \le 1/v(\lambda)$ and
$z_1 \ge 1/v(1)$, hence
\begin{equation}
(2e)^{-1} \le y_0 \le (1+\lambda)^{-1} \ .
\label{v27}\end{equation}

\subsection{The functions $H_\pm$}

\label{Hfcns}
We define
\begin{equation}
H_\pm(w) = \psi(\pm e^{\beta w}),\ \ \ \ 
\beta = \log (1/\lambda)\ ,\ \ \ \ 
\wh H_\pm = 1-H_\pm\ .
\label{h1}\end{equation}
$H_+$ is holomorphic in rhe cut strip
\begin{equation}
\Delta_+(\lambda) = \{w \in \bC\ :\ |\Im w| <\pi/\beta\} \setminus (2 + \bR_+).
\label{h1.1}\end{equation}
It maps points in $\bC_\pm$ into
$\bC_\mp$. It is decreasing on the reals, tends to 1 at $-\infty$,
and vanishes at 0. $H_-$ is holomorphic in the cut strip
\begin{equation}
\Delta_-(\lambda) = \{w \in \bC\ :\ |\Im w| <\pi/\beta\} \setminus (1+\bR_+),
\label{h1.2}\end{equation}
maps points in $\bC_\pm$ into $\bC_\pm$,
is increasing on the reals and tends to 1 at $-\infty$.
They satisfy
\begin{eqnarray}
H_\pm(w) &=& V(H_\mp(w-1)) = W(H_\pm(w-2))\ ,\nonumber \\
\wh H_\pm(w) &=& \wh V(\wh H_\mp(w-1)) = \wh W(\wh H_\pm(w-2))\ .
\label{h2}\end{eqnarray}
Moreover
\begin{equation}
{H_\pm''(w) \over H_\pm'(w)} = {\wh H_\pm''(w) \over \wh H_\pm'(w)} =
\beta \left ( 1 + {z \psi''(z) \over \psi'(z)} \right ),\ \ \ \ 
z = \pm e^{\beta w}\ .
\label{h3}\end{equation}
Since (for $0<\nu\le 1$) $\psi$ is anti-Herglotzian in
$\bC_+\cup\bC_-\cup (-1/\lambda,\ 1/\lambda^2)$,
the inequalities (\ref{AA.0}) imply, for $0 < z = e^{\beta w}< 1/\lambda$,
i.e. for all $w \in (-\infty,\ 1)$,
\begin{equation}
{H_+''(w) \over H_+'(w)} \ge
\beta \left ( 1 - {2\lambda z \over 1+\lambda z} \right )
\ge 0 \ .
\label{h4}\end{equation}
For $0 < -z = e^{\beta w}< 1/\lambda$, i.e. again 
for all $w \in (-\infty,\ 1)$,
we find similarly that 
\begin{equation}
{H_-''(w) \over H_-'(w)} \ge
\beta \left ( 1 + {2\lambda^2 z \over 1-\lambda^2 z} \right )
\ge 0\ .
\label{h5}\end{equation}
In other words, $H_-$ and $\wh H_+$ are increasing and convex,
$H_+$ is decreasing and concave.
In particular, for $w < 0$, using (\ref{v17}),
\begin{eqnarray}
2 \wh H'_+(w) &\ge& \wh H_+(w) - \wh H_+(w-2) \nonumber \\
&=&
\wh H_+(w) - \wh W^{-1}(\wh H_+(w))
\ge c \wh H_+(w)^3\ .
\label{h6}\end{eqnarray}
Integrating this with the initial condition $\wh H_+(0) =1$ gives
\begin{eqnarray}
\wh H_+(w) &\le& (1-cw)^{-1/2},\nonumber \\
H_+(w) &\ge& 1- (1-cw)^{-1/2}\ \ \ \ \ \forall w \in \bR_-\ \ \ 
(c = 1/9).
\label{h7}\end{eqnarray}
Similarly, defining $\unl H_-(w) = H_-(w)-1$, 
recalling that $H_-(0) = \psi(-1)$, $H_-(-1) = \psi(-\lambda)$, 
we obtain, using (\ref{v25}),
\begin{displaymath}
\unl H'_-(w) \ge k \unl H_-(w)^3/2\ ,
\end{displaymath}
\begin{equation}
H'_-(w) \ge  k(H_-(w)-1)^3/2
\ \ \ \ \forall w \in \bR_-\ \ \ 
(k = 1/(6e + 3(2e -1)^2))\ .
\label{h8}\end{equation}
We will need a lower bound for $H'_-(w)/H_-(w)$ in the interval
$w \in [-1,\ 0]$. This is provided by the lower bound 
$H_-(-1) = \psi(-\lambda) \ge 1+\lambda$, and by (\ref{h8}) :
\begin{eqnarray}
{H'_-(w) \over H_-(w)} &\ge& 
{k (H_-(w)-1)^3 \over 2 H_-(w)} \nonumber \\
&\ge&
{k (H_-(-1)-1)^3 \over 2 H_-(-1)}
\ge {k \lambda^3 \over 2(1+\lambda)}\ \ \ \ \ 
\forall w \in [-1,\ 0]\ .
\label{h9}\end{eqnarray}

\subsection{Lower bound on $\tau$}

\label{taubd}
Recall that the function $\vhi$ satisfies
\begin{eqnarray}
\vhi(z) &=& \lambda^{\nu-1}\, \check u^p(\lambda z),\ \ \ \ 
\forall z \in \bC_+\cup\bC_-\cup(-1/\lambda,\ 1/\lambda^2),\ \ \ \ 
\vhi(1) = 1,\nonumber \\
\vhi'(1) &=& \tau^\nu\ 
= \lambda^\nu \prod_{j=0}^{p-1} \check u'(\zeta_j),
\ \ \ \lambda \le \zeta_j = \check u^j(\lambda) \le \lambda^{1-\nu}
\ .
\label{t1}\end{eqnarray}
Let $T(w) = e^{\beta\,w}$, $\beta = \log(1/\lambda)$. Then the function
\begin{equation}
X = T^{-1} \circ \vhi \circ T
\label{t2}\end{equation}
is given by
\begin{eqnarray}
X(w) &=& -\nu +1 + Y^p(w-1) \ \ \ \ \forall w \in (-\infty,\ 2),
\nonumber \\
Y(w) &=& T^{-1} \circ \check u \circ T(w) \nonumber \\
&=& {\log y_0 \over \log(1/\tau)} +\nu -1 + 
{1  \over \log(1/\tau)}\,\log H_-(w) \ \ \ \ \forall w \in (-\infty,\ 1)\ .
\label{t3}\end{eqnarray}
It satisfies $X(0) = 0$ and
\begin{equation}
X'(0) = \tau^\nu = \prod_{j=0}^{p-1} Y'(w_j)
= \prod_{j=0}^{p-1} {1\over \log(1/\tau)}
{H_-'(w_j) \over H_-(w_j)}\ ,
\label{t4}\end{equation}
where
\begin{equation}
-1 \le w_j = {\log \zeta_j \over \log(1/\lambda)} 
\le \nu-1\ .
\label{t5}\end{equation}
Hence by (\ref{h9}),
\begin{equation}
\tau^{\nu/p-3/r}\,\log(1/\tau) \ge {k \over 4}\ ,\ \ \ \ \ 
k = 1/(6e + 3(2e -1)^2)\ .
\label{t6}\end{equation}
When $r > 3p/\nu$, this provides a lower bound for $\tau$.
We may e.g. rewrite (\ref{t6}) as
\begin{equation}
y\,\log(1/y) \ge (\nu/p -3/r)k/4,\ \ \ \ \ \ 
y = \tau^{\nu/p-3/r}\ .
\label{t7}\end{equation}

\subsection{Limiting fixed points}

\label{limfp}
The preceding subsections have shown that, for any solution, the associated 
functions have the following properties:

\begin{description}
\item{(1)} The function $V$ is holomorphic and anti-Herglotzian
in $\bC_+\cup\bC_-\cup(0,\ \zeta_{\max})$, where 
$\zeta_{\max} \ge \tau^{\nu-2} \ge 8^{(2-\nu)/\nu}$. It satisfies
$V(1) =1$ and $V'(1) = -1/\lambda$.

\item{(2)} The function $W = V\circ V$ is holomorphic and Herglotzian
in $\bC_+\cup\bC_-\cup(0,\ \alpha)$, where 
$(1+\lambda) \le \alpha = V^{-1}(0) \le e^\lambda$.

\item{(3)} The function $H_+$ is holomorphic in the cut strip
$\Delta_+(\lambda)$ (see (\ref{h1.1})),
maps points in $\bC_\pm$ into $\bC_\mp$, vanishes at 0, and 
satisfies the bound (\ref{h7}).

\item{(4)} The function $H_-$ is holomorphic in the cut strip
$\Delta_-(\lambda)$ (see (\ref{h1.2})),
maps points in $\bC_\pm$ into $\bC_\pm$, and satisfies 
$H_-(-1) = \alpha$ and the bounds (\ref{h8}) and (\ref{h9}).

\item{(5)} $\tau = \lambda^r$ is bounded above by $\tau \le 8^{-1/\nu}$.
For sufficiently large $r$, its is bounded below by (\ref{t6}), and for
all $r$ by $\lambda_0(p,\ r,\ \nu)^r$ (see (\ref{4.23.13})).

\item{(6)} $y_0 = z_1^r$ satisfies (\ref{v27}).
\end{description}

\noindent
As a consequence every infinite sequence of solutions, 
with fixed $\nu$ and $p$, such that $r \rightarrow \infty$,
contains an infinite subsequence such that $\tau$ and
$y_0$ have limits in $(0,\ 1)$, and that the functions $V$, $W$,
$H_\pm$ tend, uniformly over compact sets,
to non-constant functions, holomorphic in cut planes.
Meanwhile, $\lambda$ and $z_1 > \lambda^\nu$ tend to 1 (see (\ref{p.3a})),
$\psi$ and $u$ tend
to 1, uniformly over compact subsets of 
$\bC_+\cup\bC_-\cup(-1,\ 1)$ (see (\ref{h7})). However the functions
\begin{equation}
S_\pm (\zeta) = U(\pm \zeta^{1/r})
= y_0\tau^{1-\nu} H_\pm \left({\log \zeta \over \log(1/\tau)}\right )
\label{l1}\end{equation}
have non-trivial limits and obey the functional equation:
\begin{equation}
S_\pm (\zeta) =  {1\over \tau^\nu}
S_+(\tau^{\nu-1}\, S_-^{p-1}\circ S_\mp(\tau \zeta))\ .
\label{l2}\end{equation}

\appendix
\section{Appendix 1. Proof of the inequality (\ref{4.23.2.1})}
This inequality is equivalent to
\begin{equation}
(1-x^2)\,\log((1+x^2)(1+x)^2) + 4x^2\,\log (x) > 0 \ \ \ \ 
\forall x \in (0,\ 1)\ ,
\label{A1.1}\end{equation}
or to
\begin{equation}
f_1(x) - 4 xf_2(x)  > 0 \ \ \ \ 
\forall x \in (0,\ 1)\ ,
\label{A1.2}\end{equation}
where
\begin{equation}
f_1(x) = \log((1+x^2)(1+x)^2),\ \ \ \ 
f_2(x) = {x\,\log(1/x) \over 1-x^2}\ .
\label{A1.3}\end{equation}
The derivative $f'_2(x)$ has the sign of
\begin{equation}
-2\log(x) - 2\left ({1-x^2 \over 1+x^2}\right ) 
= -\log(t) - {4\over 1+t} +2, \ \ \ t=x^2\ .
\end{equation}
The last expression vanishes at 1 and has negative derivative in $t$
on $(0,\ 1)$. Hence 
$f_2$ is increasing on $(0,\ 1)$. It tends to $1/2$ as $x$ tends to 1,
so that $f_2 < 1/2$ on $(0,\ 1)$. It now suffices to prove that
$f_1(x) -2 x > 0 $ for all $x \in (0,\ 1)$.
This quantity vanishes for $x=0$, and 
\begin{equation}
f'_1(x) - 2 \ \ =\ \ {2x^2 (1-x) \over (1+x^2)(1+x)} > 0  \ \ \ \ 
\forall x \in (0,\ 1)\ .
\end{equation}

\end{document}